\documentclass[useAMS,usenatbib]{mn2e}

\usepackage[psamsfonts]{amssymb}
\usepackage[dvips]{graphicx}
\usepackage{amsmath,alltt}
\usepackage{multirow}
\usepackage{multicol}

\usepackage{rotating}
\usepackage{lscape}
\title[Stellar dynamics in gas]{Stellar dynamics in gas:  The role of gas damping}
\author[Leigh N. W. C., Mastrobuono Battisti A., Perets H. B. \& B\"{o}ker T.]{Nathan W. C. Leigh$^{1,2}$, 
Alessandra Mastrobuono Battisti$^{3}$, Hagai B. Perets$^{3,4}$, 
\newauthor
Torsten B\"{o}ker$^{5}$
\thanks{E-mail: nleigh@ualberta.ca (NL), alessandra.mastrobuono@gmail.com (AM), hperets@ph.technion.ac.il (HP), 
tboeker@rssd.esa.int (TB)}\\
$^{1}$Department of Physics, University of Alberta, CCIS 4-183, Edmonton, AB T6G 2E1, Canada \\
$^{2}$Department of Astrophysics, American Museum of Natural History, Central Park West and 79th Street, New York, NY 10024 \\
$^{3}$Physics Department, Technion: Israel Institute of Technology, Haifa, Israel 32000 \\
$^{4}$Deloro Fellow \\
$^{5}$European Space Agency, Space Science Department, Keplerlaan 1, 2200 AG Noordwijk, The Netherlands}
\begin{document}

\pagerange{\pageref{firstpage}--\pageref{lastpage}} \pubyear{2011}

\maketitle

\label{firstpage}

\begin{abstract}
In this paper, we consider how gas damping affects the dynamical evolution of gas-embedded star clusters.  
Using a simple three-component (i.e. one gas and two stellar components) model, we compare the 
rates of mass segregation due to two-body relaxation, accretion from the interstellar medium, and gas 
dynamical friction in both the supersonic and subsonic regimes.  
Using observational data in the literature, we apply our analytic predictions to two
different astrophysical environments, namely galactic nuclei and young open star clusters.  
Our analytic results are then tested using numerical simulations 
performed with the NBSymple code, modified by an additional deceleration term to model the damping 
effects of the gas.  

The results of our simulations are in reasonable agreement with our analytic 
predictions, and demonstrate that gas damping can significantly accelerate the rate of mass segregation.  
A stable state of approximate energy equilibrium cannot be achieved 
in our model if gas damping is present, even if Spitzer's Criterion is satisfied.  This instability 
drives the continued dynamical decoupling and subsequent ejection (and/or collisions) of the more 
massive population.  Unlike two-body 
relaxation, gas damping causes overall cluster contraction, reducing both the core and half-mass 
radii.  If the cluster is mass segregated (and/or the gas density is highest at the cluster centre), the 
latter contracts faster than the former, accelerating the rate of core collapse.
\end{abstract}


\section{Introduction} \label{intro}

The advances made in telescope resolution over the last few decades have revealed 
the properties of a number of interesting astrophysical environments that 
contain both gas and stars in significant quantities.  For example, the 
properties of many young star-forming regions throughout our Galaxy have now been 
catalogued 
%.  In many cases, the stellar mass functions in these gas-rich star clusters 
%have been measured down to below the hydrogen burning limit 
using the unprecedented 
spatial resolution of the Hubble Space Telescope (HST), and large ground-based telescopes 
assisted by adaptive optics \citep[e.g.][]{wang11,phan-bao11,demarchi11,dario12}.  The 
gas being used to form stars in these regions is typically dense and cold, with 
densities and temperatures on the order of $\sim 10^2-10^6$ M$_{\odot}$ pc$^{-3}$ and 
$\sim 10$ K (temperatures can reach $\sim 30$ K during the later stages of star formation) 
\citep[e.g.][]{lada03,mckee07}, respectively.  The nuclei of spiral galaxies 
have also been imaged at high resolution, revealing not only that 
significant quantities of molecular gas are present, but also that star formation 
is on-going in these dense stellar environments \citep[e.g.][]{boker03,schinnerer06}.  
The implications of the presence of gas in galactic nuclei, both with respect to 
the formation of nuclear star clusters and the growth of super-massive black holes (SMBHs) 
are not yet fully understood \citep[e.g.][]{hopkins11,gabor13,kormendy13}.  
%For example, early on in the life-times of early-type galaxies, some of the 
%gas goes into fueling the growth of a central SMBH, a nuclear star cluster, or 
%both.  How exactly this occurs is not known.  Somehow, the process must trace the 
%origin of the M-$\sigma$ relation for SMBHs.  Whether or not SMBH growth can 
%be attributed directly to gas accretion, the M-$\sigma$ relation offers important 
%constraints for the conditions in, and evolution of, primordial gas-rich galactic nuclei.
%
%\textbf{Brief paragraph about SMBH formation in galactic nuclei.  M-$\sigma$ relation?}
%
%By contrast, the interstellar medium (ISM) in older stellar populations, in 
%particular globular clusters (GCs), is noteably sparse, containing at most a few tens of 
%solar masses of gas \citep[e.g.][]{taylor75,frank76,priestley11}.  However, 
Additionally, recent evidence suggests that globular clusters (GCs) underwent 
prolonged star formation early on in their lifetimes (see \citet{gratton12} for a recent 
review).  That is, gas was present in significant quantities for the first $\sim 10^8$ years, 
albeit perhaps intermittently \citep{conroy11,conroy12}.  This evidence comes 
in the form of multiple stellar populations identified in the colour-magnitude 
diagram \citep[e.g.][]{piotto07}, as well as curious abundance anomalies that cannot be 
explained by a single burst of star formation \citep[e.g.][]{osborn71,gratton01}.  
%This suggests 
%that gas was present in primordial GCs in significant quantities for the first $\sim 10^8$ years, 
%albeit perhaps intermittently \citep{conroy11,conroy12}.  
%, since this corresponds roughly to the derived age-spreads 
%between multiple populations \citep[e.g.][]{conroy11,conroy12}.  This indirectly 
%implies that, 
Thus, at least until the second generation has formed, stars from the first 
generation must have been orbiting within a gas-rich environment.  

On the theoretical front, most studies conducted to date considered one of two extremes.  The 
first has its focus purely on 
the stellar dynamics (see \citet{spitzer87} and \citet{heggie03} for detailed reviews), long 
after the gas has been converted to stars and/or ejected from the cluster.  The focus of the 
second extreme is often on the very early stages of star formation 
\citep[e.g.][]{krumholz05b,kirk11,krumholz11a,krumholz11b,offner11,krumholz12,kirk14}, 
when gas is first being converted into stars.  Little theoretical work has been done connecting these 
two extremes, when both gas and stars co-exist in significant quantities, particularly in 
massive clusters and for more than a few tens of Myr.  

\citet{bonnell97} first studied gas accretion 
onto small clusters of young stars using a three-dimensional SPH code.  The authors find 
that non-uniform or differential accretion can produce a realistic mass spectrum, even when the 
initial stellar masses are uniform, and tends to form massive stars at the cluster centre where 
the gas density is highest \citep[e.g.][]{bonnell98,bonnell01}.  Later, \citet{bate03} performed 
a larger numerical simulation to resolve the fragmentation process down to the opacity limit.  
The authors find that the star formation process is highly chaotic and dynamic, and that 
the observed statistical properties of stars are a natural consequence of a dynamical 
environment.  These results were later expanded upon by \citet{bate09}, \citet{bate12} and 
\citet{maschberger10} to include more massive clusters composed of $\sim$ a few thousand stars.  
Other authors placed their focus on the evolving properties of the gas \citep{offner09a}, the impact 
of different initial conditions \citep{girichidis12a} or the evolution 
of potential and kinetic energy throughout the cloud \citep{girichidis12b}, and showed that clusters 
tend to be born in a subvirial state \citep[e.g.][]{allison09b}.  Both 
the gas velocity dispersion and temperature can be surprisingly high, especially if radiative 
feedback is included in the simulations \citep[e.g.][]{offner09a,offner09b}.  Along with 
large-scale magnetic fields \citep[e.g.][]{lee14}, this can dramatically 
reduce not only the overall accretion rate, but also the final number of stars that form.  A 
number of $N$-body simulations have also been performed to study the effects of gas expulsion or 
dispersal \citep[e.g.][]{marks08,moeckel12}, sub-cluster merging \citep[e.g.][]{moeckel09b}, 
primordial mass segregation \citep[e.g.][]{marks08,moeckel09a}, and even techniques to quantify 
the degree of mass segregation \citep{allison09a,parker12,parker14}.

In purely stellar systems (i.e. gas-free), two-body relaxation is the dominant physical mechanism 
driving the evolution of star clusters for most of their lifetime 
\citep[e.g.][]{henon60,henon73,spitzer87,heggie03,gieles11}.  
Long-range gravitational interactions tend to push the cluster toward a state of energy 
equipartition in which all objects have comparable kinetic energies.\footnote{In reality, such an 
idealized state is never fully achieved in a cluster with a realistic mass spectrum or potential; 
see \citet{trenti13} for more details.}  Consequently, the velocities of the most massive 
objects decrease, and they accumulate in the central regions of the cluster.  Similarly, the 
velocities of the lowest mass objects increase, and they are
subsequently dispersed to wider orbits.  This mechanism is called 
mass segregation, and has been observed in both young and old 
clusters \citep[e.g.][]{lada03,vonhippel98,demarchi10}.

In the presence of gas, other effects could also contribute to the deceleration of a massive 
test particle.  For example, mass accretion reduces the accretor velocity due to conservation 
of momentum.  This was first argued by \citet{bondi44} and \citet{bondi52} in the supersonic and 
subsonic limits, respectively.  
%derived the rate of mass accretion by assuming that the gas 
%temperature is small throughout the cluster, and hence that the forces due to gas pressure are negligible 
%compared to the force of gravity.  Later, \citet{bondi52} derived the solution in the opposite 
%limit, namely when gravitational forces are small compared to those due to gas pressure.  He 
%argued that the solutions found for these two limits should bracket the range of plausible 
%accretion rates for all gas temperatures.  Importantly, the velocity of the accretor was 
%shown to be reduced in both limits due to conservation of momentum.  
%More recently, 
%\citet{leigh13} expanded on the two-component model of \citet{spitzer69}, 
%and derived the time-scale for mass accretion to cause a gas-embedded star cluster to 
%become mass segregated.  The authors argued that for large cluster masses 
%(M $\gtrsim 10^4-10^5$ M$_{\odot}$) and high gas densities (n $\gtrsim 10^5$ cm$^{-3}$), 
%accretion from the interstellar medium (ISM) could significantly accelerate the rate of 
%mass segregation in a primordial star cluster.  
%
Several authors have also explored the effects of gas dynamical friction, particularly in 
the steady-state supersonic regime \citep[e.g.][]{dokuchaev64,ruderman71,rephaeli80,ostriker99}, 
although it is most effective when the perturber velocity is approximately 
equal to the sound speed of the surrounding gaseous medium.  
%For example, \citet{ostriker99} 
%argued that the timescale for gas dynamical friction could be shorter than the 
%corresponding timescale for stellar dynamical friction, given certain reasonable 
%assumptions.  
More recently, \citet{lee11} and \citet{lee13} showed that the drag force should be precisely 
equal to $\dot{m}$v, where $\dot{m}$ is the rate of mass accretion and v is the 
accretor's velocity relative to the gas, in both the subsonic and supersonic regimes.  
Consequently, the authors argued that damping due to gas dynamical friction cannot be 
separated from that due to accretion, and that these processes instead represent different 
components of the same underlying damping mechanism.
%This is perhaps not surprising, since the drag 
%induced by both accretion and gas dynamical friction are due to the same basic physics, 
%namely the gravitational pull of a moving object on the surrounding gaseous medium.  In 
%the limit of very small velocities, the motion of the gas surrounding 
%the test particle achieves approximate spherical symmetry (ignoring accretion 
%disks), and both drag forces should converge to zero.

In this paper, we address the question:  When does the presence of gas in a 
stellar system significantly affect the stars' dynamics?  To answer this question, we will 
adopt a three-component model to calculate and compare three timescales for a massive 
test particle orbiting within the cluster potential to become mass segregated.  The damping 
mechanisms we consider are two-body relaxation, 
accretion from the interstellar medium (ISM) and gas dynamical friction.  To first 
order, this will allow us to constrain the parameter space in the cluster mass-gas 
density plane for which each of the different damping mechanisms dominates the deceleration 
of the test particle.  We then apply our results to observed data taken from the 
literature for two different types of astrophysical environments, namely the galactic 
nuclei of late-type spirals and young star-forming regions, and determine the 
dominant damping mechanism operating in each.  

We compare all three timescales in both the subsonic and supersonic regimes in Section~\ref{analytic1},  
%The computational $N$-body models we use to test 
%our analytic predictions are presented in Section~\ref{compute}.  
%Our analytic results are presented in Section~\ref{results}, 
and apply our analytic model to the available 
observational data.  The modified $N$-body simulations used to model the effects of gas damping 
are presented in Section~\ref{comp2}, and compared to the predictions of our analytic model, the 
details of which are provided in an Appendix.  Finally, in Section~\ref{discussion}, we 
summarize our key results, and discuss the significance of our results for different 
astrophysical environments, in particular galactic nuclei and young star-forming regions.
%where 
%either accretion from the ISM or gas dynamical friction could dominate the rate 
%of mass segregation.  
%We conclude in Section~\ref{summary}.

\section{Analytic model} \label{analytic1}

In this section, we present the results of our analytic model, the details of which are 
presented in Appendix A, along with a discussion of our model assumptions.  Using our model, we 
derive the approximate 
time for a massive test particle to become mass segregated -- i.e. reach an orbit within the cluster 
that is approximately consistent with energy equipartition or, more accurately, energy equilibrium 
\citep{trenti13} -- in 
a spherical, self-gravitating, pressure-supported, stellar system 
embedded in a gaseous medium.  This is done by evaluating the deceleration induced 
on the particle by, and hence the timescale for, each of the relevant damping mechanisms to reduce 
the speed of the test particle from $\sigma$ to $\sigma\sqrt{\bar{m}/m_{\rm 1}}$, 
%\footnote{\textbf{As we will 
%show, the final velocity plays a negligible role in deciding the mass segregation times due to gas 
%dynamical friction and accretion.  Hence, the actual shape of the gravitational potential plays only a 
%negligible role in our calculations of these timescales, and we do not concern ourselves with the 
%technical details concerning energy equipartition versus energy equilibrium.}}, 
where $\sigma$ is the root-mean-square velocity, $\bar{m}$ is 
the average particle mass and m$_{\rm 1}$ is the mass of the test particle.  
To first-order, we ignore the shape of the gravitational potential, and assume that gas damping will 
act to smoothly decrease the root-mean-square speeds of the stars over time.\footnote{As we will 
show, the final velocity plays a negligible role in deciding the mass segregation times due to gas 
dynamical friction and accretion.}  
The damping mechanisms we consider are two-body relaxation, gas dynamical friction and accretion 
from the ISM.  
Using our analytic timescales, we further solve for the parameter space in the cluster mass-gas 
density plane for which each of these mechanisms dominates the deceleration of a massive test 
particle.

%\textbf{To first-order, the stellar orbits are viewed as pendulum-like, such that gas damping will 
%act to smoothly decrease the root-mean-square speeds of the stars over time.}

In principle, our analytic model is suitable to a number of astrophysical scenarios, including a 
massive star 
orbiting within a gas-embedded star cluster, a primordial globular cluster orbiting within 
the natal Milky Way, or even galaxies orbiting within galaxy clusters.  For the remainder of 
this paper, however, we present our model within the context of a gas-embedded star 
cluster in order to keep the discussion as simple and direct as possible, and to 
facilitate comparisons to the rich body of observational
data and theoretical models for resolved star clusters available in the
literature.\footnote{Note that, since we focus on the star cluster case in this paper, we do 
not consider 
stellar dynamical friction, which is more appropriate for larger systems where the assumption 
is valid that the massive body undergoing stellar dynamical friction moves 
in an infinite sea of low-mass particles.}  
%
%We use a three-component model to derive the timescale for each damping mechanism to reduce 
%the speed of our test particle from $\sigma$ to $\sigma\sqrt{\bar{m}/m_{\rm 1}}$.  This 
%signifies energy equipartition, and hence mass segregation.  
%The model and timescales are presented in Appendix A, along with a discussion of our model assumptions.  
The results of our model are 
applied to different astrophysical environments for which the relevant observational measurements are 
available in the literature, specifically young star-forming regions and galactic nuclei.

\subsection{Timescales} \label{times}

\begin{figure*}
\centering
\resizebox{0.46\hsize}{!}{\rotatebox{0}{\includegraphics{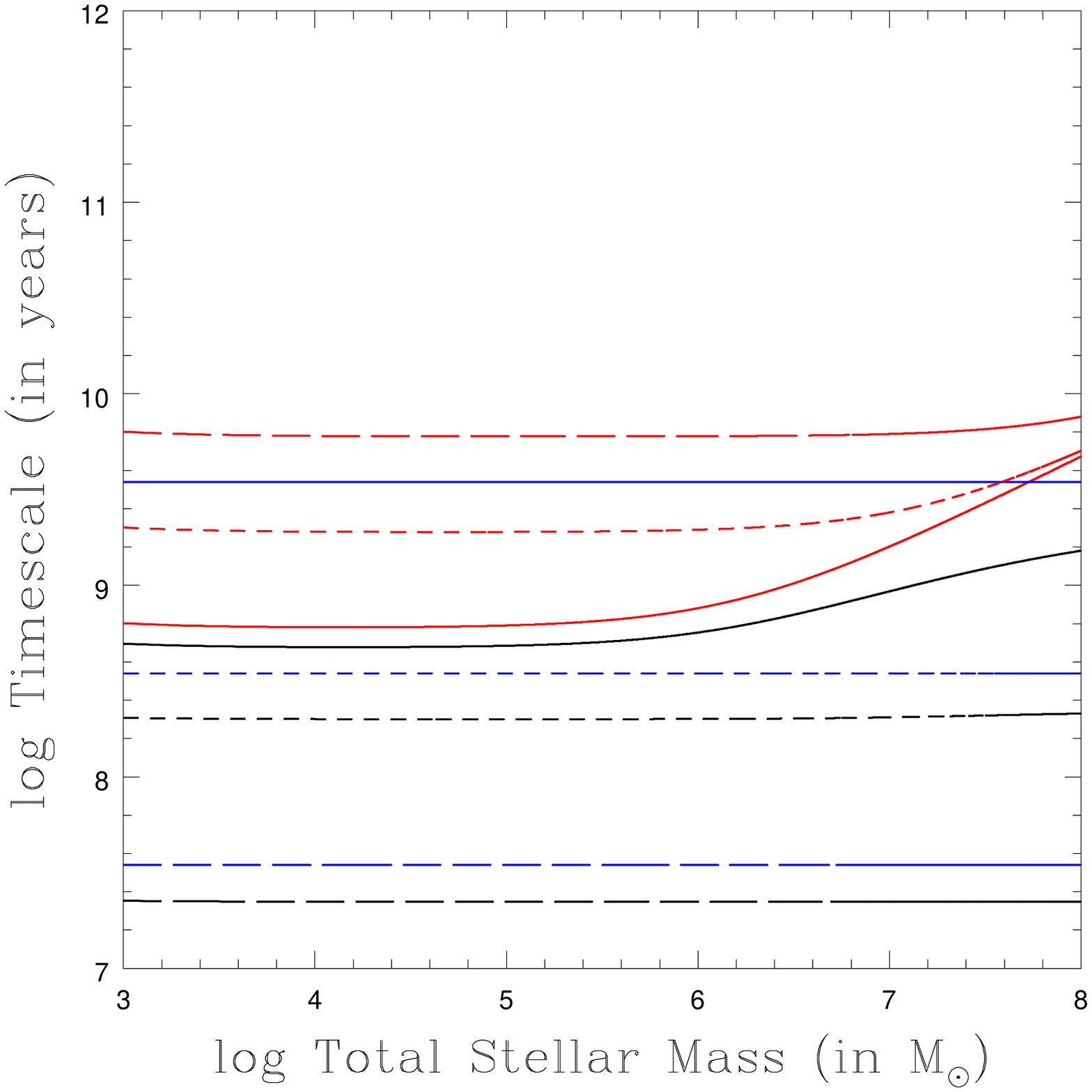}}}
\resizebox{0.46\hsize}{!}{\rotatebox{0}{\includegraphics{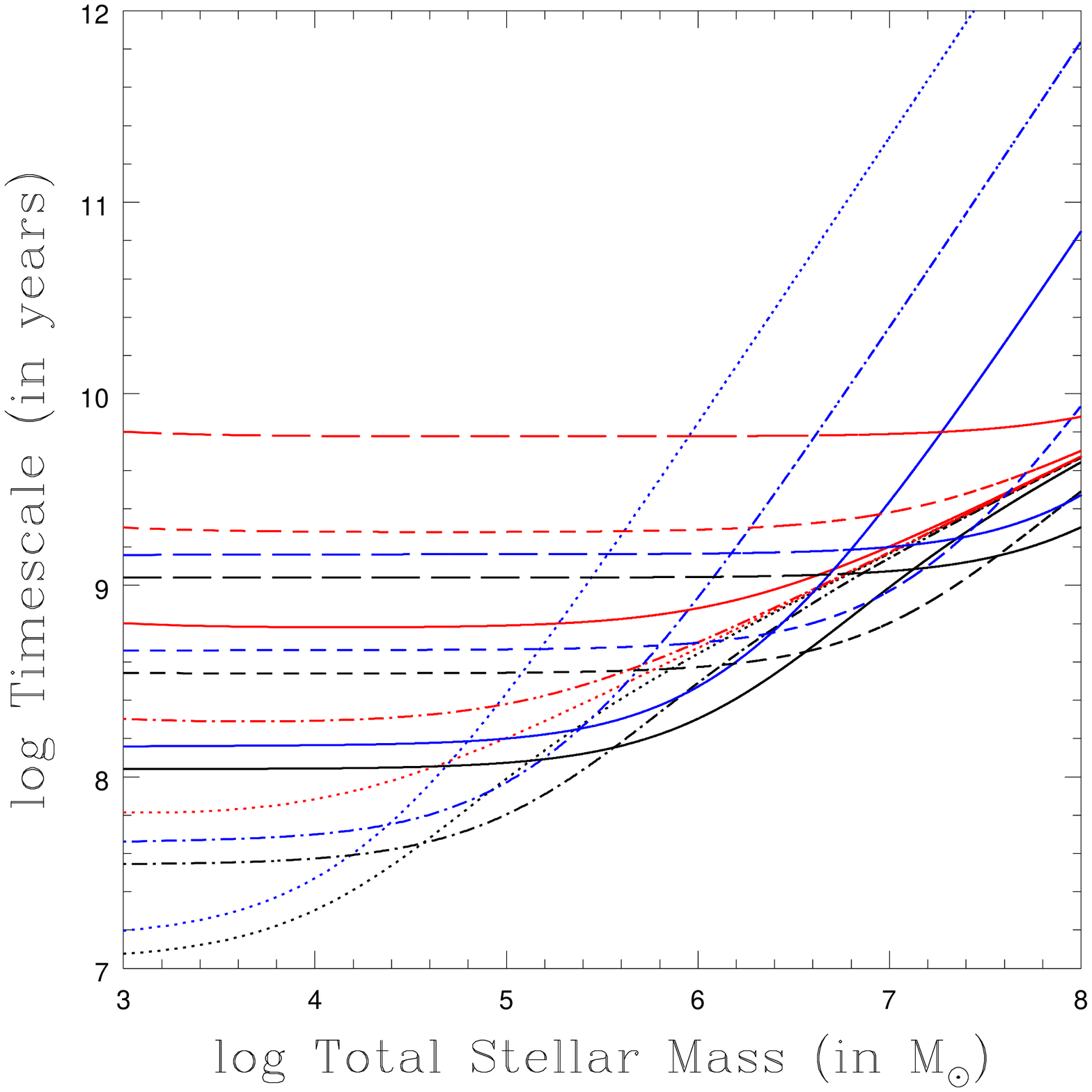}}}
\caption[Timescales for two-body relaxation and gas dynamical friction as a function of the total
stellar mass for both the subsonic and supersonic cases]{Timescales for two-body relaxation and 
gas dynamical friction are 
shown as a function of the total stellar mass assuming subsonic motion with
$c_{\rm s} = 10$ km/s (left panel), or supersonic motion with $c_{\rm s} = 0.1$ km/s (right panel).
The mass of the test particle is taken to be 32.55 M$_{\odot}$.  The red lines correspond to the
two-body relaxation timescales, 
%the blue lines to the accretion time-scales, 
the blue lines to the gas dynamical friction
timescales, and the black lines to the cumulative or total timescales calculated from the
sum of the rates of all three mechanisms.  The dotted, dash-dotted, solid, short-dashed, and 
long-dashed lines correspond to average gas densities of 10 cm$^{-3}$, 10$^2$ cm$^{-3}$, 
10$^3$ cm$^{-3}$, 10$^4$ cm$^{-3}$ and 10$^5$ cm$^{-3}$, respectively.
\label{fig:fig1}}
\end{figure*}

We show in Figure~\ref{fig:fig1} the timescales for gas dynamical friction and two-body relaxation as 
a function of the 
total stellar mass (the mass in gas is not included on the x-axis) for both the subsonic 
(left panel) and supersonic (right panel) cases.  We adopt r$_{\rm h} = 20$ pc, and do not plot the 
timescale for accretion to 
avoid over-populating Figure~\ref{fig:fig1}.  Instead, we show the cumulative timescale, calculated 
from the sum of the rates of all three damping mechanisms (i.e. including accretion).  Recall that, 
for a given test particle mass, the timescale for accretion to operate is directly proportional 
to the timescale for gas dynamical friction.

%\subsubsection{Subsonic} \label{subsonic1}

In the subsonic case, shown in the left panel of Figure~\ref{fig:fig1}, 
the timescale required for gas accretion to operate is 
always longer than the gas dynamical friction (blue lines) timescale by a factor $\sim 2.5$.  
However, both timescales are considerably shorter than the corresponding two-body relaxation time 
at high gas densities.  Specifically, at a gas density 10$^3$ cm$^{-3}$ (shown by the solid lines in 
Figure~\ref{fig:fig1}) two-body relaxation dominates over gas dynamical friction provided the total 
stellar mass $\lesssim 10^{7.7}$ M$_{\odot}$.  
At higher gas densities (shown by the short- and long-dashed lines in 
Figure~\ref{fig:fig1}), gas dynamical friction always dominates the rate of 
mass segregation, and two-body relaxation always plays a negligible role independent of the 
total stellar mass.

In the supersonic case, shown in the right panel of Figure~\ref{fig:fig1}, 
the timescale for gas dynamical friction (blue lines) to operate is always 
shorter than the accretion timescale (not shown in Figure~\ref{fig:fig1}) by a factor $\sim 10$.  
The gas dynamical friction timescale tends to be shorter than the two-body relaxation timescale (red lines)  
for low total stellar masses, but becomes longer than the two-body relaxation timescale for high 
total stellar mass.  This transition shifts to higher total stellar masses as the gas density increases.  
We note as well that for larger r$_{\rm h}$, gas dynamical friction and accretion can 
dominate over two-body relaxation at lower gas densities, but assuming a fixed total stellar mass.  
The perhaps counter-intuitive result that the two-body relaxation timescale 
is \textit{shorter} than the timescales for gas accretion and gas dynamical friction at large 
total cluster masses can be understood as follows.  The timescales for both accretion and 
gas dynamical friction scale as v$^3$, where v is the velocity of the test particle with respect to the 
gas.  A higher gas density implies a higher total gas mass, which translates into higher stellar 
velocities via the virial theorem.  

\subsection{Identifying the dominant damping mechanism} \label{mass-density}

\begin{figure*}
\centering
\resizebox{0.46\hsize}{!}{\rotatebox{0}{\includegraphics{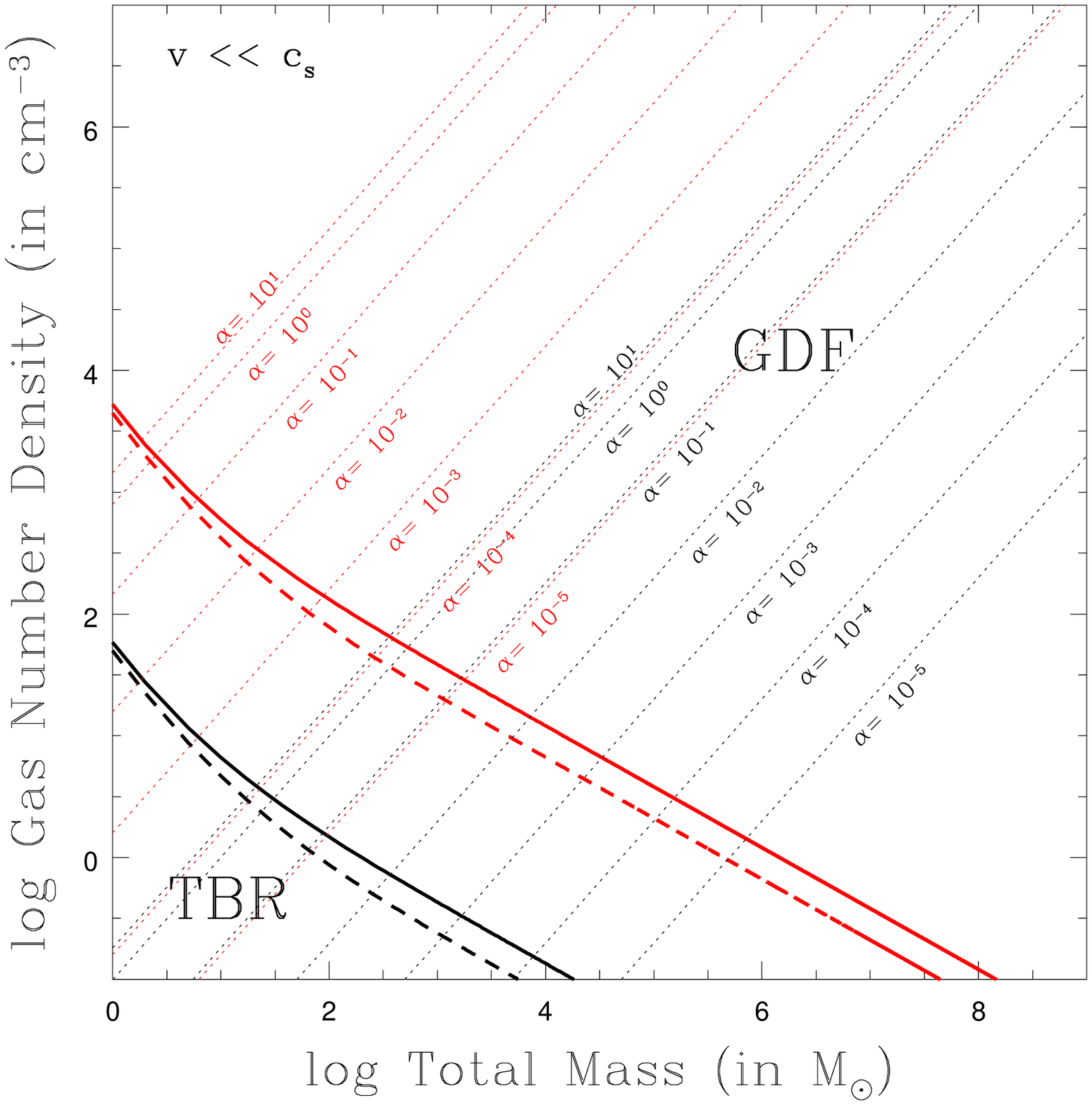}}}
\resizebox{0.46\hsize}{!}{\rotatebox{0}{\includegraphics{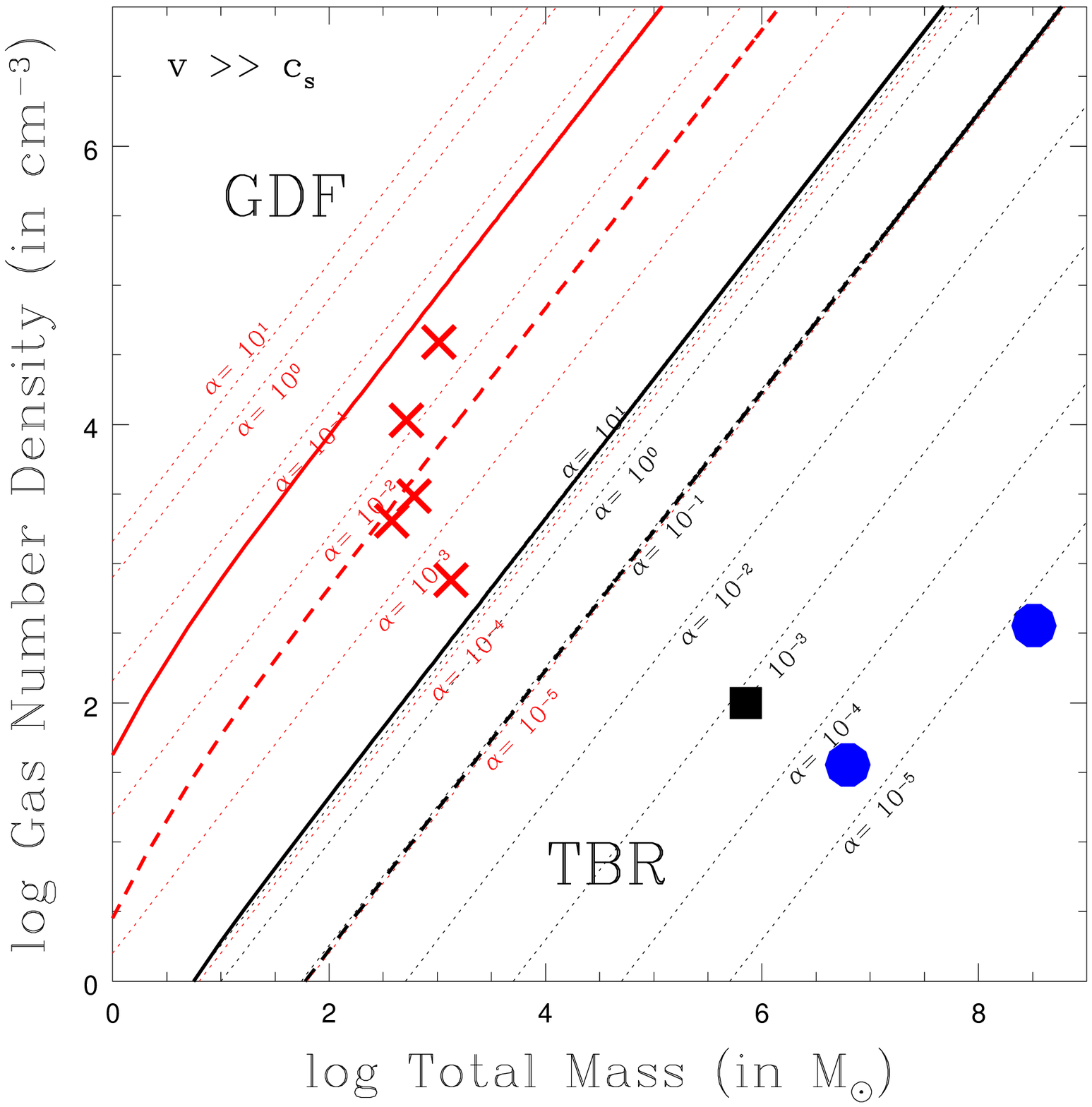}}}
\caption[Parameter space in the cluster mass-gas density plane for which each damping mechanism
dominates]{The parameter space in the cluster mass-gas density-plane
for which each damping mechanism dominates the deceleration of a massive test particle.  This is shown for 
both subsonic (left panel) and 
supersonic (right panel) motion.  The regions dominated by two-body relaxation and gas dynamical 
friction are indicated by ``TBR'' and ``GDF'', respectively.  The total cluster mass, shown on the 
x-axis, is calculated as the sum of the total mass in gas and 
stars.  The mass of the test particle is taken to be 32.55 M$_{\odot}$.  The black lines 
adopt a half-mass radius for the cluster of r$_{\rm h} = 20$ pc, whereas the red lines adopt 
r$_{\rm h} = 1$ pc.  The solid lines 
%Right now, plots assume r$_{\rm h} =$ 20 pc.
correspond to $\tau_{\rm rh} = \tau_{\rm acc}$, and the long-dashed lines to
$\tau_{\rm rh} = \tau_{\rm df}$.  The dotted lines correspond to constant gas mass fractions 
$\alpha =$ M$_{\rm g}$/M$_{\rm s}$.  The blue circles correspond to the nuclei
of the late-type spiral galaxies NGC 6946 \citep{schinnerer06} and IC 342 \citep{schinnerer03}.
The black square is the Milky Way Galactic Centre \citep{launhardt02}.  The red crosses
correspond to observed data taken from \citet{lada03} for young
open clusters or associations.
\label{fig:fig2}}
\end{figure*}

Next, we show in Figure~\ref{fig:fig2}, for both the subsonic (left panel) and supersonic (right 
panel) cases, the parameter space in the cluster mass-gas density plane for which each of the 
different damping mechanisms dominates the rate of deceleration of a massive test particle.  The 
regions dominated by two-body relaxation and gas dynamical friction are indicated by ``TBR'' and 
``GDF'', respectively.  The solid and dashed lines in Figure~\ref{fig:fig2} correspond 
to the relations between the total (gas and stars) cluster mass and average gas density 
obtained by equating each combination of the derived timescales.  These relations are 
$\tau_{\rm df} = \tau_{\rm rh}$ (solid lines) and $\tau_{\rm acc} = \tau_{\rm rh}$ (dashed 
lines).  These lines divide the parameter space for which gas dynamical friction and accretion 
each dominate over two-body relaxation.  We exclude the relation $\tau_{\rm df} = \tau_{\rm acc}$ 
since, as shown in the previous section, the rate of 
deceleration due to gas dynamical friction is always greater than the rate for gas 
accretion.  The dotted lines show the cluster mass-gas density relations for constant gas mass 
fractions $\alpha =$ M$_{\rm g}$/M$_{\rm s}$.

\subsubsection{Subsonic} \label{sub3}

As shown in the left panel of Figure~\ref{fig:fig2}, two-body relaxation can dominate at low total 
cluster masses in the subsonic regime, 
provided the gas density is also low.  At sufficiently large total cluster masses, however, gas 
damping dominates over two-body relaxation independent of the gas density.  
For instance, for a total 
cluster mass $\sim 10^6$ M$_{\odot}$ and half-mass radius r$_{\rm h} = 1$ pc, gas damping 
dominates for all gas densities.  
If the half-mass radius r$_{\rm h}$ increases, the lines shift to a lower gas density at fixed 
cluster mass, 
and the distance between the lines increases slightly, with increasing particle mass.
%CHECK THE ABOVE!!!

\subsubsection{Supersonic} \label{sup3}

In the supersonic regime, very high gas densities are needed in massive clusters for accretion and 
gas dynamical friction to supersede 
two-body relaxation as the dominant damping mechanism.  This is illustrated in the right panel of 
Figure~\ref{fig:fig2} by the positive slope of the lines.  For large cluster masses and low gas densities 
(i.e. for small gas mass fractions), 
two-body relaxation dominates over both accretion and gas dynamical friction.  At low cluster 
masses, both accretion and gas dynamical friction dominate independent of the gas density, provided 
r$_{\rm h} \gtrsim 15$ pc.  
%For larger cluster masses, however, two-body relaxation dominates at low gas densities.  
%For large r$_{\rm h}$ 
%(r$_{\rm h} \gtrsim 20$), increasingly high 
%gas densities are needed for accretion and gas dynamical friction to supersede two-body relaxation 
%as the dominant damping mechanism.  This is illustrated in the right panel of Figure~\ref{fig:fig2} 
%by the positive 
%slope of the lines.  
As r$_{\rm h}$ decreases, the y-intercept of the lines shifts to higher gas 
densities at fixed cluster mass.  Both 
the slopes and y-intercepts of the lines are roughly insensitive to our choice for the mass of 
the test particle.

\subsection{Application to observational data} \label{app}

In order to apply our results to observed gas-rich clusters, we take data from the 
literature to obtain total cluster masses and 
average gas densities.  This is done for two different types of astrophysical environments, 
namely galactic nuclei and young star-forming regions.  Specifically, we apply our model to 
the galactic nuclei of the Milky Way (black squares in Figure~\ref{fig:fig2}), NGC 6946 and 
IC 342 (blue circles) \citep{launhardt02,schinnerer06,schinnerer03}, as well as to the young
star-forming regions Mon R2, NGC 1333, NGC 2024, NGC 2068, NGC 2071 (red crosses) 
\citep{lada03}.  The observed data for these nuclear and open clusters are provided in 
Table~\ref{table:obs}, and the results of applying our model to these data are shown in 
Figure~\ref{fig:fig2}.  The location of each 
point in the mass-density-plane tells us which damping mechanism is currently dominating 
in that environment.  \textit{In order to see this, the red crosses should be compared 
to the red lines, and the black squares and blue circle should be compared to the black lines.}  We 
include the observed data points only in the right panel of Figure~\ref{fig:fig2} since, as we 
explain below, we expect the stellar motions to be in the supersonic regime in all environments 
considered here.

%\begin{minipage}{16cm}
\begin{center}
\begin{table*}
%{|c|c|c|c|}
\caption{Observational parameters for each cluster/nucleus}
%\centering
\begin{tabular}{|l|c|c|c|}
\hline
\hline
\multicolumn{1}{c}{Cluster/Nucleus}  &  log Total Mass$^{\rm a}$ (in M$_{\odot}$)  &  log Gas Density (in cm$^{-3}$)  &  Size (in pc)  \\
\hline
Mon R2$^{\rm b}$    &  3.13  &  2.88  &    1.85  \\
NGC 1333            &  3.01  &  4.59  &    0.49  \\
NGC 2024            &  2.79  &  3.49  &    0.88  \\
NGC 2068            &  2.58  &  3.31  &    0.86  \\
NGC 2071            &  2.71  &  4.03  &    0.59  \\
IC 342$^{\rm c}$    &  6.79  &  1.55  &   30.00  \\
NGC 6946$^{\rm d}$  &  8.51  &  2.55  &   60.00  \\
MW$^{\rm e}$        &  5.85  &  2.00  &    1.25  \\
\hline
\hline
\multicolumn{4}{l}{$^{\rm a}$Including both the stellar and gas mass.}\\
\multicolumn{4}{l}{$^{\rm b}$All open cluster data is taken from \citet{lada03}.}\\
\multicolumn{4}{l}{$^{\rm c}$Data for the late-type spiral galaxy NGC 6946 is taken from \citep{schinnerer06},} \\
\multicolumn{4}{l}{and 30 pc is the limiting radius considered in the study.}\\
\multicolumn{4}{l}{$^{\rm d}$Data for the late-type spiral galaxy IC 342 is taken from \citep{schinnerer03},} \\
\multicolumn{4}{l}{and 60 pc is the limiting radius considered in the study.}\\
\multicolumn{4}{l}{$^{\rm e}$Data for the Milky Way Galactic Centre is taken from \citep{launhardt02} and} \\
\multicolumn{4}{l}{\citet{merritt13}, and corresponds only to the inner 1.25 pc.}\\
%\end{tabular}
\label{table:obs}
%\end{table*}
\end{tabular}
\end{table*}
\end{center}
%\end{minipage}
%\end{center}

The gas in the inner 1.25 pc of the Galactic Centre is hot and ionized.  However, we still 
expect the stellar motions to most likely be in the supersonic regime due to the large stellar velocities.  
This may also be the 
case for IC 342, which is thought to be similar to the Milky Way Galactic Centre.  The right panel of 
Figure~\ref{fig:fig2} suggests that two-body relaxation dominates in these environments if the motion is 
supersonic.  Even if the stars are in the subsonic regime, the left panel of Figure~\ref{fig:fig2} 
suggests that two-body relaxation should still dominate in these regions.  Importantly, this remains 
the case if the black lines in Figure~\ref{fig:fig2} are shifted to correspond directly to the observed 
half-mass radii for these nuclear star clusters, both in the supersonic and subsonic regimes.  

The gas in the remaining environments shown in Table~\ref{table:obs} is predominantly molecular
hydrogen.  Hence, we expect the stellar motions to be mainly supersonic.  In this regime, 
the right panel of Figure~\ref{fig:fig2} suggests that 
two-body relaxation dominates in all but two young open clusters.  These are NGC 1333 and NGC 2071, 
which have the highest gas densities.  Importantly, however, 
our model adopts r$_{\rm h} = 1$ pc for the red lines, which is slightly too 
large for these two low-mass open clusters (see Table~\ref{table:obs}).  For smaller r$_{\rm h}$, the 
y-intercepts of the lines in Figure~\ref{fig:fig2} shift upward to higher gas densities.  For NGC 1333 
and NGC 2071, the end result is that the rates of mass segregation due to two-body relaxation and gas 
dynamical friction are roughly equal.  Thus, we conclude that for the open clusters considered here, 
the rate of mass segregation due to two-body relaxation tends to be comparable to, but slightly exceeds, 
the rates from gas dynamical friction and accretion.  

\section{Computational models} \label{comp2}

In this section, we present and discuss the results of our computational $N$-body simulations modified 
to include the effects of gas damping (but not the gas potential itself).  

\subsection{The models} \label{models}

We wish to test the analytic theory presented in the previous section.  This is done to evaluate the effects of gas
damping on the overall cluster evolution, and the degree to which gas damping can accelerate the rate of mass
segregation - i.e. the rate at which the most massive members in a self-gravitating system end up at the bottom of
the potential well.  To do this, we use computational $N$-body models for star cluster evolution performed using an
adapted version of the NBSymple code \citep{capuzzo-dolcetta11} to simulate mass segregation in a gas-embedded star
cluster.  We consider only the steady-state supersonic limit in our simulations, since the gas in most astrophysical
cases of interest is typically cold and molecular, with a correspondingly low sound speed.

%We are unable to explicitly include a gas component in our $N$-body models, due to the considerable
%additional computational expense.  Thus, to 
To model the effects of the gas, we introduce an additional
deceleration on each $N$-body particle equal to that due to gas dynamical friction, given in
Equation~\ref{eqn:fric-force}.  This is used to reduce the velocity of every $N$-body particle at
each time-step.  We consider only the deceleration induced by gas dynamical friction in our computational
models, since a quick comparison of the timescales derived in the preceding section reveals that gas dynamical
friction is more effective than gas accretion in the supersonic regime (for the range of particle velocities and
masses considered here).  We note that, neglecting the dependence on the test particle mass, the timescales for
accretion and gas dynamical friction are equivalent to within a dimensionless constant, given by the ratio between
Equation~\ref{eqn:tau-acc4} and Equation~\ref{eqn:tau-df3}.  Thus, our chosen form for the deceleration on a massive
test particle is representative of gas damping in general to within a dimensionless constant.

For all models, we assume a gas composed entirely of hydrogen molecules (i.e. m$_{\rm 3} =$ 3.3 $\times$ 10$^{-27}$ kg),
with a gas density n $=$ 4 $\times$ 10$^3$ cm$^{-3}$.  The sound speed is chosen to be very low
c$_{\rm s} \ll \sigma$ to ensure that the motion is always in the steady-state supersonic regime.  
This means that our chosen form for the deceleration induced by gas damping given in 
Equation~\ref{eqn:fric-force} is always applicable.  All models assume
stellar masses m$_{\rm 1} =$ 32.55 M$_{\odot}$ and m$_{\rm 2} =$ 3.255 M$_{\odot}$.  These masses are chosen to 
yield the desired total cluster masses for our simulations (see below), 
however they are in general representative of stellar masses observed in young massive star-forming 
regions.  Stellar evolution does not occur in any of our simulations, so that the stellar masses are time-independent.

We consider two different 
cases.  Case 1 adheres to Spitzer's Criterion, with the total mass in species 1 being much less than the
total mass in species 2.  If Spitzer's Criterion is satisfied, then a stable state of energy equilibrium 
(e.g. energy equipartition) should 
be achievable between both mass species.  If not, the heavier species decouples dynamically from the lighter 
species, interacting primarily with other members of the heavier species in the central cluster regions.  Case 1 
is tailored for comparison to our analytic model.  Case 2 assumes that 
the total number of stars is split equally between the two mass-components.  In this case, we do not compare
our analytic calculations to the results of our simulations, since this deviates from the assumptions of our
analytic model.  Dividing the total number of stars equally between both mass
species allows us to construct meaningful surface density profiles for each of them individually.  This
facilitates quantifying the effects of gas accretion and gas dynamical friction on the overall cluster
structure.  We also follow the evolution of the cluster considered in Case 2 to core collapse.
%Stellar evolution does not occur in any of our simulations, so that the stellar masses are time-independent.

For Case 1, the total number of stars belonging to species 1 and 2 are, respectively, N$_{\rm 1} = 100$ and
N$_{\rm 2} = 15260$.  This gives a total cluster mass 5.29 $\times$ 10$^4$ M$_{\odot}$, and an average
stellar mass of $\bar{m} =$ 3.45 M$_{\odot}$.  The initial half-mass radius is 3.9 pc before the cluster 
experiences an intial phase of expansion, after which it is 6.1 pc (at t $=$ 3 Myr).  
The cluster orbits within the Galactic potential on a mildly eccentric orbit with
e $=$ 0.05, and a distance at perigalacticon of 9.7 kpc.  From Equation~\ref{eqn:t-rh}, the initial
two-body relaxation time is 1.8 $\times$ 10$^8$ years.

For Case 2, the total cluster mass 2.75 $\times$ 10$^5$ M$_{\odot}$ and the total number of stars is
N$_{\rm s} =$ 15360, giving an average stellar mass $\bar{m} =$ 17.9 M$_{\odot}$.  The initial core and
half-mass radii are 2 pc and 4.05 pc, respectively, and the cluster is modeled assuming a
King profile with W$_{\rm 0} = 5$, initially.  The cluster orbits within the Galactic potential
assuming the same orbit as in \citet{odenkirchen03}.
From Equation~\ref{eqn:t-rh}, the initial half-mass relaxation timescale is 4.1 $\times$ 10$^7$ years.

\subsection{Comparison to analytic predictions} \label{model-results}

%In this section, we compare our analytic predictions to the results of our computational models, and 
%discuss the implications for the early phases of star cluster evolution.  
%
%\subsubsection{Case 1:  Spitzer's criterion} \label{case1}

\begin{figure*}
\centering
\resizebox{0.46\hsize}{!}{\rotatebox{0}{\includegraphics{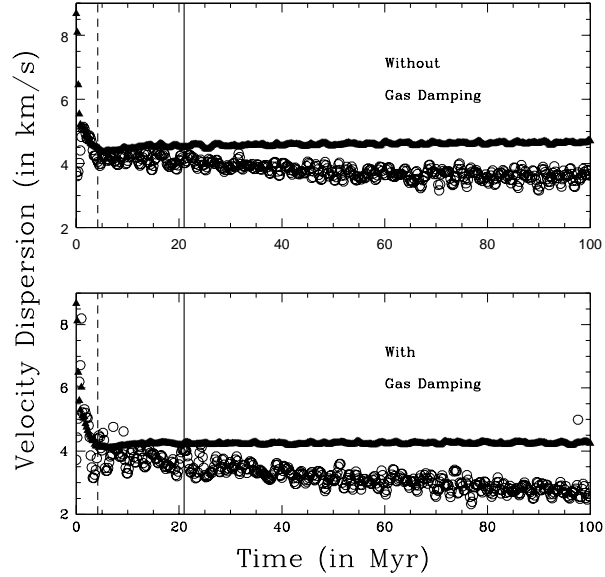}}}
\caption[Time evolution of the root-mean-square speeds of both mass species in Case 1 models with and
without gas damping]{The time evolution of the root-mean-square speeds for the low- (filled triangles) and 
high-mass (open circles) species are shown, both with (bottom inset) and without (top inset) 
gas damping.  Only stars within 5 pc of the cluster centre are included in calculating the root-mean-square 
speeds, to avoid including stars that are ejected from the cluster.  The dashed and solid lines show our 
analytic predictions for the timescale for mass 
segregation due to gas dynamical friction and two-body relaxation, respectively, beginning at 
t $=$ 3 Myr, which corresponds roughly to the time at which the initial episode of mild violent 
relaxation ceases, and hence t $=$ 0 in our analytic model.    
\label{fig:fig3}}
\end{figure*}

We begin by comparing the results of our simulations to the predictions of our analytic model, which predicts 
the time at which the heavier species reaches a mean-square speed
$\sim$ $\sigma\sqrt{\bar{m}/m_{\rm 1}}$ from an initial mean-square speed $\sigma$.  For our models, 
this implies that the mean-square speed of the heavier species (species 1) must fall by a factor 
$\sim 3$ (i.e. from $\sim 3.8$ kms$^{-1}$ to $\sim 1.3$ kms$^{-1}$).  
%
%Figure~\ref{fig:fig3} shows the distribution of stellar velocities for the 
%low- (top insets) and high-mass (bottom insets) species, both with (right insets) and without (left 
%insets) gas damping.  The velocities are shown at t $=$ 0 Myr (black) and t $=$ 500 Myr
%(red).  
Figure~\ref{fig:fig3} shows the time evolution of the root-mean-square speeds of both the light (filled triangles) 
and heavy (open circles) species, both with (bottom inset) and without (top inset) gas damping.  Initially, 
the clusters expand in our simulations, undergoing an episode of mild violent relaxation.\footnote{This episode of 
violent relaxation is not accounted for in our analytic model.  The initial conditions of our model are only met 
after this phase of evolution ceases.}  This phase ends at t $\sim$ 3 Myr, at which point steady-state is achieved, 
and the system begins to evolve toward a state of energy equilibrium.  This corresponds to t $=$ 0 in 
our analytic model.

The onset of energy equilibrium 
is depicted in Figure~\ref{fig:fig3}, since the root-mean-square speed of the heavier species begins to 
fall below that of the lighter species.  We observe significantly more scatter in the time evolution of the 
root-mean-square speed of the heavier species, particularly when gas damping is present.  This is due mainly 
to the small number of particles for species 1, but also in part due to strong gravitational interactions that occur 
between members of species 1 upon segregating into the core.  Due to this scatter, the exact time at which energy 
equilibrium and mass segregation occur is ambiguous in Figure~\ref{fig:fig3}.  Even more problematic, members of 
species 1 quickly segregate into the core where the velocity dispersion is at its highest, and this also contributes 
to increasing the mass 
in the core, which further increases the central velocity dispersion.  Hence, exact energy \textit{equipartition} 
does not occur in the simulations without gas damping, since the root-mean-square speed of the heavier species never 
drops as much as energy equipartition initially predicts.  Specifically, the root-mean-square speed of the heavier
species only falls by at most a factor $\sim 1.5$ relative to the lighter species after several tens of Myr, at which
point the system stabilizes and energy \textit{equilibrium} is reached.  This is less than predicted by energy 
equipartition, as discussed in more detail in 
\citet{trenti13}.  Only with gas damping does the root-mean-square speed of the heavier species continue to 
fall below that of the lighter species beyond what is done by two-body relaxation alone.  Regardless, it is clear from 
Figure~\ref{fig:fig3} that gas damping accelerates the rate at which energy equilibrium, and hence mass segregation, 
occurs by a factor $\sim 2$ relative to two-body relaxation alone for the model assumptions adopted here.

The predictions of our analytic model are \textit{qualitatively} borne out by the simulations, since 
gas damping accelerates the rate of mass segregation.  Quantitatively, however, our analytic estimates 
for the mass segregation timescales under-predict the true timescales by a factor $\sim 5-10$, both without 
and especially with gas damping.  Specifically, 
from Equation~\ref{eqn:tau-df3}, mass segregation should begin occurring after 1.2 Myr if only gas dynamical 
friction is operating (shown by the dashed lines in Figure~\ref{fig:fig3}), whereas the half-mass two-body 
relaxation time (solid lines) for species 1 
is 18 Myr, as given by Equation~\ref{eqn:tau-relax} (and ignoring the mass in gas, which is 
not accounted for in our computational models).  Although some members of species 1 do indeed begin to segregate 
into the core in as a little as a few Myr, the process continues for another few tens of Myr before most members 
of species are in the core.  Thus, our analytic timescales correspond better to the \textit{onset} of mass 
segregation, as opposed to its termination.  Other contributing factors to the discrepany between our analytic timescale and the 
results of our simulations include the steep dependence of the efficiency of gas damping on the particle velocity, 
and our adopted estimate of the velocity dispersion (using the approximation in \citet{binney87}) which is a slight
underestimate for members of species 1 throughout the course of our simulations.  
%Qualitatively, our analytic model predicts an increase in the rate of 
%mass segregation with gas damping, as observed in our simulations.  Quantitatively, however, our analytic 
%timescales are too short by a factor $\sim 5-10$.  For the gas damping timescale, part of this discrepancy 
%can be attributed to the steep dependence on the particle velocity, combined with the fact that our 
%estimate of the central velocity dispersion (using the approximation in \citet{binney87}) is a slight 
%underestimate for some members of species 1.  
%Additionally, 
%we note that in the simulations without gas damping, the root-mean-square speed of the heavier 
%species only falls by at most a factor $\sim 1.5$ relative to the lighter species after several tens of Myr, at which 
%point the system stabilizes.  This is less than predicted by energy equipartition, as demonstrated in 
%\citet{trenti13}.  Only with gas damping does the root-mean-square speed of the heavier species continue to 
%fall below that of the lighter species beyond what is done by two-body relaxation alone.

\begin{figure*}
\centering
\resizebox{0.46\hsize}{!}{\rotatebox{0}{\includegraphics{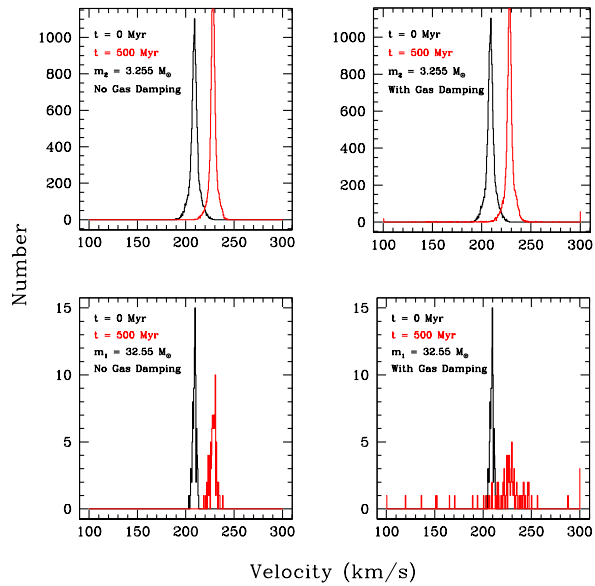}}}
\caption[Distribution of stellar velocities in Case 1 models for both mass species with and 
without gas damping]{The distribution of stellar velocities are shown for the low- (top insets) 
and high-mass (bottom insets) species, both with (right insets) and without (left insets) 
gas damping.  The velocities are shown at t $=$ 0 Myr (black) and t $=$ 500 Myr 
(red).  
%The cluster is on a mildly eccentric orbit, with apogalacticon and perigalacticon speeds 
%of $\sim$ 205 and 230 km/s, respectively.  This accounts for most of the offset between the black 
%and red distributions, since the latter corresponds roughly to apogalacticon and the former to shortly 
%after perigalacticon.  The smaller peaks at $\sim$ 205 km/s and 215 km/s correspond to tidal 
%tails, which contain approximately half the initial total cluster mass.
\label{fig:fig4}}
\end{figure*}

Figure~\ref{fig:fig4} shows snapshots in time of the distribution of stellar velocities for the
low- (top insets) and high-mass (bottom insets) species, both with (right insets) and without (left
insets) gas damping.  The velocities are shown at t $=$ 0 Myr (black) and t $=$ 500 Myr
(red).  The cluster is on a mildly eccentric orbit, with apogalacticon and perigalacticon speeds
of $\sim$ 205 and 230 km/s, respectively.  This accounts for most of the offset between the black
and red distributions, since the latter corresponds roughly to apogalacticon and the former to shortly
after perigalacticon.  The smaller peaks at $\sim$ 205 km/s and 215 km/s correspond to tidal
tails, which contain approximately half the initial total cluster mass.

Interestingly, in the simulations with gas damping, the cluster is unable to remain in equilibrium, 
and the heavier species reaches a sufficiently high central density to decouple dynamically from the 
lighter species.  This 
leads to strong gravitational interactions between members of the heavier species, and ultimately 
their continual ejection from the cluster over time.  With gas damping, a large 
fraction of species 1 has been ejected from the cluster within $< 500$ Myr, whereas no members of species 1 
have been ejected in the simulation without gas damping.  This illustrates that gas 
damping can significantly accelerate 
the dynamical ejection of a massive sub-population in clusters, or even stimulate such a population 
to decouple dynamically and undergo this phase of ejections when it otherwise would remain stable.

\begin{figure*}
\centering
\resizebox{0.46\hsize}{!}{\rotatebox{0}{\includegraphics{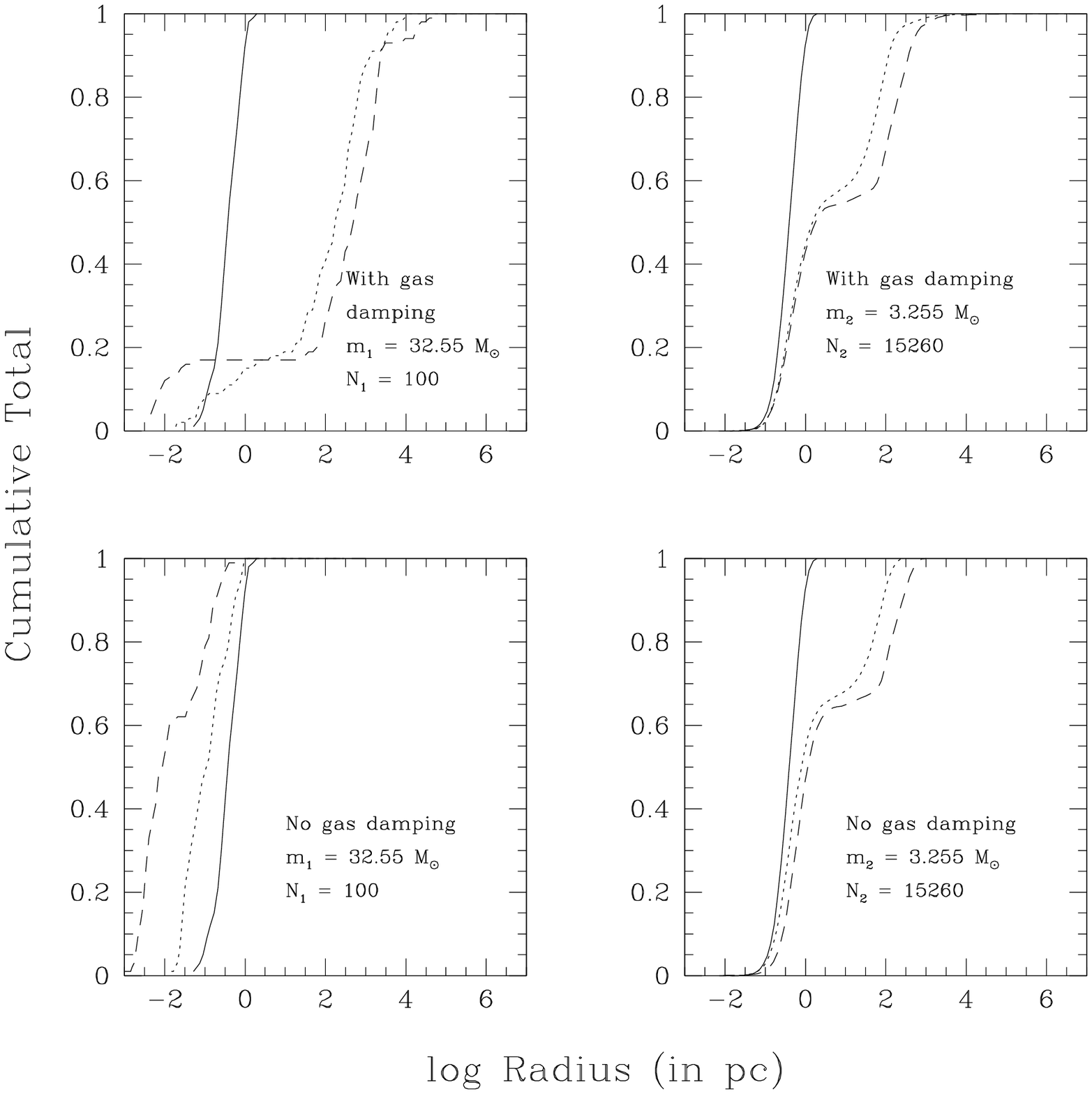}}}
\caption[Cumulative radial density profiles for the high- and low-mass species, both with and without 
gas damping, for N$_{\rm 1} =$ 100 and N$_{\rm 2} =$ 15260]{The cumulative radial density profiles are 
shown for the low- (bottom insets) 
and high-mass (top insets) species, both with (left insets) and without (right insets) 
gas damping.  The radial profiles are shown at t $=$ 0 Myr (solid lines), t $=$ 100 Myr (dotted lines) 
and t $=$ 500 Myr (dashed lines).  The number objects belonging to the heavier species is 
N$_{\rm 1} =$ 100, whereas for the lighter species N$_{\rm 2} =$ 15260.  Thus, the total 
mass in species 1 is much less than the total mass in species 2, and Spitzer's Criterion is 
satisfied.
\label{fig:fig5}}
\end{figure*}

This last point is further illustrated in Figure~\ref{fig:fig5}, which shows the cumulative 
radial density profiles for the high- (left insets) and low-mass (right insets) species, 
both with (top insets) and without (bottom insets) gas dynamical friction.  For the high-mass 
species, the bottom left inset shows that mass segregation has occurred in $\ll$ 100 Myr, 
with all members of species 1 being confined to the central cluster regions.  This configuration 
remains stable in the case without gas damping for the next 500 Myr.  With gas damping, however, 
the heavier species decouples dynamically in $\ll$ 100 Myr, and strong gravitational interactions 
between members of species 1 have ejected a large fraction of the heavier population 
to large cluster radii, albeit many remain on bound orbits and return to the core within a few 
Myr.  This process of evaporation through strong encounters \citep{henon69} results in the complete 
ejection of $\gtrsim 80\%$ of the heavier population by 500 Myr.  

Approximately half the cluster mass begins to form tidal tails well 
within 100 Myr, as shown in the cumulative radial density profiles for the lighter species in 
Figure~\ref{fig:fig5}.  This 
is because the cluster is initially tidally over-filling.  Note that the evolution is significantly 
more rapid in the simulations with gas damping.

%SHOULD WE INTEGRATE OVER A MAXWELLIAN VELOCITY DISTRIBUTION TO OBTAIN TIME-SCALES FOR AN
%ENTIRE POPULATION OF TEST PARTICLES, FOR BETTER DIRECT COMPARISON TO THE SIMULATIONS?

%\subsubsection{Case 2:  Equal population sizes} \label{case2}

%\subsection{Implications for cluster evolution and structure} \label{implications}

\begin{figure*}
\centering
\resizebox{0.46\hsize}{!}{\rotatebox{0}{\includegraphics{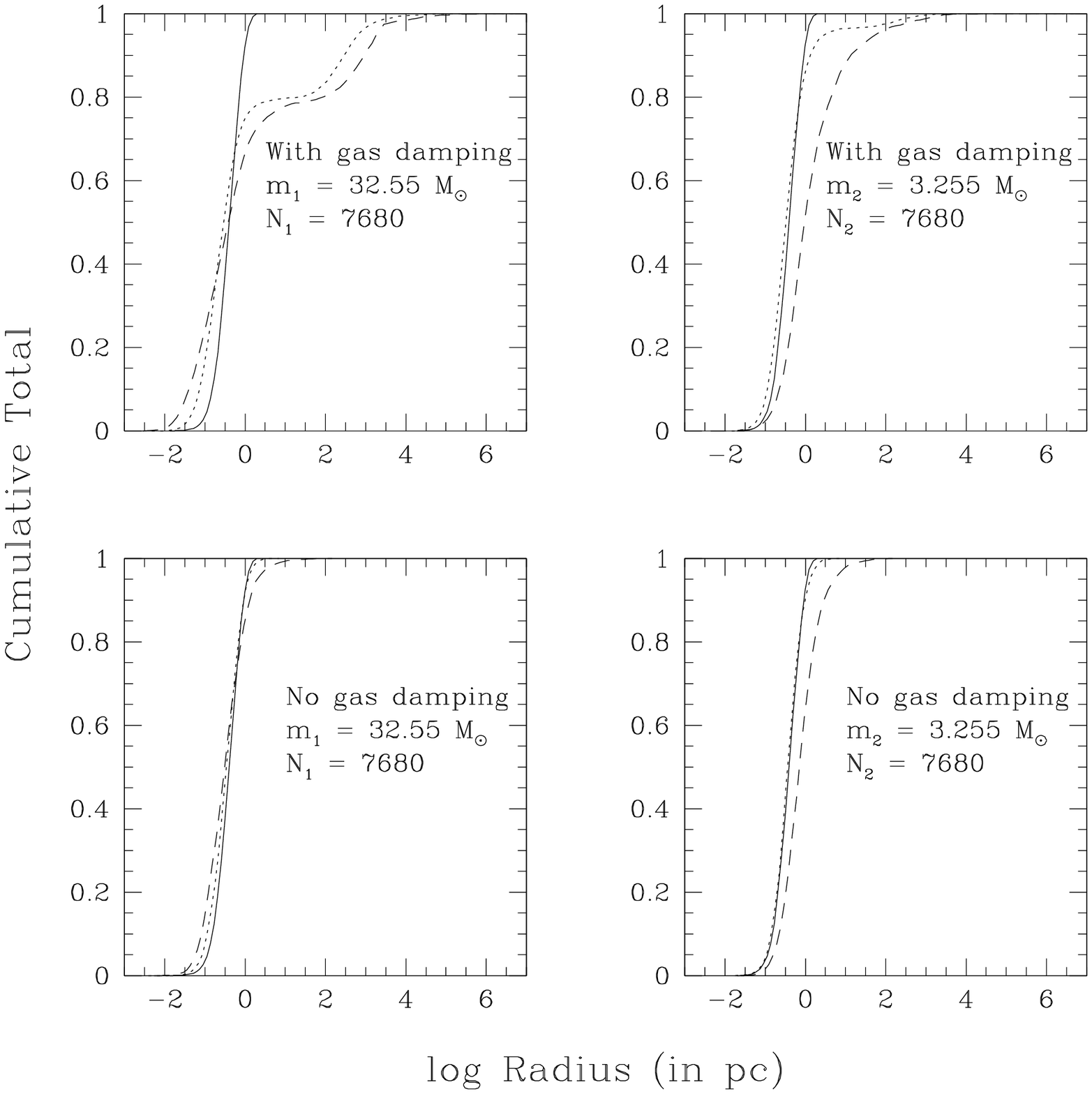}}}
\caption[Cumulative radial density profiles for the high- and low-mass species, both with and without
gas damping, for N$_{\rm 1} =$ N$_{\rm 2} =$ 7680]{The cumulative radial density profiles are shown 
for the low- (bottom insets) 
and high-mass (top insets) species, both with (left insets) and without (right insets)
gas damping.  The radial profiles are shown at t $=$ 0 Myr (solid lines), t $=$ 100 Myr (dotted lines)
and t $=$ 500 Myr (dashed lines).  The number of objects belonging to each mass species is equal, 
with N$_{\rm 1} =$ N$_{\rm 2} =$ 7680.
\label{fig:fig6}}
\end{figure*}

%In this section, we use the results of our computational models to discuss the effects 
%of gas damping for the overall cluster structure.  To this end, 
Next, we present the results of our simulations for which the initial population size 
of each species is half the total number of objects (i.e. Case 2).  This is  
shown in Figures~\ref{fig:fig6} for the supersonic case, which illustrates the 
cumulative radial density profiles for the high- (top insets) and low-mass (bottom insets) 
species, both with (top insets) and without (bottom insets) gas dynamical friction, 
at 0 Myr (solid lines), 100 Myr (dotted lines) and 500 Myr (dashed lines).  

The key result illustrated in Figure~\ref{fig:fig6} is that gas damping causes clusters to 
contract and compactify, losing stars from their outskirts to tidal tails at an accelerated rate.  
The stronger is the gas damping, the more pronounced is the effect.  This is shown by comparing 
the cumulative radial density profiles of the heavier (top left inset) and lighter (top right inset) 
species in the simulations with gas damping.  The lighter species takes considerably longer 
to undergo this contraction, and even expands on a shorter timescale in response to the loss of 
cluster mass (of the heavier species).  The heavier species, on the other hand, contracts on 
a much shorter timescale than the lighter species, since the deceleration due to gas damping is 
proportional to the square of the particle mass.

\section{Discussion} \label{discussion}

%In this paper, we study the effects of gas damping on the evolution of star clusters.  
%We consider three different damping mechanisms, namely two-body relaxation, accretion from the interstellar
%medium, and gas dynamical
%%friction.  Using a simple three-component model, we calculate the timescale for each
%mechanism to cause a massive test particle to become mass segregated in both the subsonic
%and supersonic regimes, and then test our results using numerical $N$-body simulations modified 
%to include the effects of gas damping.  
%Our analytic results are then tested using numerical $N$-body simulations
%performed with a modified version of the NBSymple code, which includes an additional
%deceleration term to model the damping effects of the gas.  The predictions of our analytic model
%are in excellent agreement with the results of our $N$-body simulations.

In the subsequent sections, we discuss the implications of our results for the evolution of 
gas-embedded star clusters, comment on the validity of our model
assumptions and offer suggestions for improvements in future work.  We further 
discuss what our results imply for different astrophysical cases of interest,
including nuclear star clusters and SMBH formation, young open clusters or associations, and 
primordial globular clusters.

%
%In this section, we begin by discussing the implications of our results for 
%the overall structure of star clusters, comment on the validity of our model 
%assumptions and offer suggestions for improvements in future work.  Next, we 
%discuss what our results imply for different astrophysical cases of interest, 
%including nuclear star clusters, young open clusters or associations, and 
%primordial globular clusters and starburst clusters.

\subsection{Mass segregation and energy equilibrium} \label{astro}
%Implications for cluster evolution in the gas-embedded phase} \label{astro}

%Here, we discuss the implications of our results for real astrophysical environments.
Here, we discuss the implications of our results for the rate of mass segregation 
in a gas-embedded star cluster within the context of energy equilibrium.  
%We go on to discuss the evolution of the overall cluster 
%structure during the gas embedded phase.  
%We discuss only the supersonic 
%limit since most astrophysical environments applicable to our model contain 
%primarily cold, molecular gas.

%\subsubsection{Mass segregation and energy equipartition} \label{mass-seg}

Within the framework of our analytic model, if initially all mass species in a cluster have the 
same root-mean-square speed, then all three 
damping mechanisms considered in this paper will push this initial state toward energy equilibrium.  
%where 0.5m$_{\rm 1}\sigma_{\rm 1} \approx$ 0.5m$_{\rm 2}\sigma_{\rm 2}$ 
%($\sigma_{\rm i}$ denotes the root-mean-square speed of species $i$, and 
%m$_{\rm 1} \gg$ m$_{\rm 2}$).  
Our simulations show that energy equipartition is never actually 
achieved, even without gas damping (see \citet{trenti13} for more details).  The analytic timescale 
presented in Appendix A and compared to the simulations in Figure~\ref{fig:fig3} nonetheless
approximately describe the timescale for mass segregation due to gas dynamical friction\footnote{More accurately, 
our analytic timescale corresponds more closely to the \textit{onset} of mass segregation due to gas 
dynamical friction.} to within an order of magnitude, since this timescale depends most sensitively 
on the \textit{initial} particle velocity, with the dependence on the \textit{final} particle 
velocity typically being negligible.

Our results illustrate that both gas dynamical friction and accretion from the ISM should
serve to accelerate the rate of mass segregation (or stratification) within gas-embedded
star clusters, operating on timescales comparable to (albeit typically smaller than) the 
corresponding timescale for two-body relaxation for typical gas densities and cluster masses.
Importantly, we do not expect either gas accretion or gas dynamical friction 
to stop operating once mass segregation is achieved.  These mechanisms will continue 
to reduce the velocities 
of more massive particles the fastest, causing the cluster to fall further away from 
energy equilibrium, and contract even further.  In particular, clusters that would 
otherwise achieve a stable state of approximate energy equilibrium will 
be forced out of this state by gas damping.  Thus, so long as gas damping continues to operate, a stable 
state of energy equilibrium cannot be achieved.  Eventually, the central density of the heavier 
particles becomes sufficiently high that they begin to undergo strong gravitational interactions, 
progressively dwindling their population size by ejecting each other from the cluster (or colliding). 

In massive clusters, gas damping could push clusters to core collapse on a shorter 
timescale than two-body relaxation alone.  %Assuming that the duration of the gas-embedded is 
%proportional to the total cluster mass, 
This could help to explain why more massive Galactic globular clusters 
are, on average, more concentrated \citep{harris96,leigh13b}, when two-body relaxation alone should 
cause the opposite.  That is, two-body relaxation causes the concentration to increase slowly 
over time, but this process operates at a rate that is inversely proportional to the cluster mass.  
For this scenario to be compatible with the observed distribution of concentrations, our analytic 
results suggest that the duration of the gas-embedded phase should be proportional to the total 
cluster mass.  This is because the additional gas mass increases the stellar velocities via the 
virial theorem, and the timescale for gas damping to decelerate a massive test particle scales 
with the cube of the particle velocity in the steady-state supersonic limit.  Thus, the 
duration of the gas-embedded phase should scale with total cluster mass as M$_{\rm clus}^{2/3}$, 
since $\tau_{\rm gd,sup} \propto$ $\sigma^3$ and $\sigma \propto$ M$_{\rm clus}^{1/2}$.  Here, 
$\tau_{\rm gd,sup}$ represents the timescale for gas damping (i.e. either gas accretion or gas 
dynamical friction) to operate in the steady-state 
supersonic limit and $\sigma$ is the stellar velocity dispersion, taken as a proxy for the typical 
relative velocity between stars and the gas.  If the duration of the gas-embedded phase does 
indeed scale with the total cluster mass, then this could predict a relation between second 
generation stars and the concentration parameter, perhaps in the form of a correlation between 
concentration and the fraction of second generation stars (see Section~\ref{gal-nuc} below).

Alternatively, assuming that the total mass in 
stars greatly exceeds the total gas mass (so that the stellar velocities do not depend strongly on 
the total gas mass, via the virial theorem), then more massive clusters could end up more 
concentrated if the gas fraction scales linearly with the total cluster mass, and the duration of 
the gas-embedded phase is independent of cluster mass.  This is because the rate of gas damping 
scales linearly with gas density, and we assume that the cluster half-mass radius r$_{\rm h}$ 
is also approximately independent of the total cluster mass \citep{harris96}.  

Importantly, we expect our main conclusions to hold if our simulations were to be re-performed with a 
realistic mass spectrum.  In particular, gas damping should still accelerate the rate of 
mass segregation, cause overall cluster contraction and even accelerate the rate of core 
collapse.  However, we caution that higher gas densities than adopted in the simulations 
performed in this paper will likely be required to achieve comparable effects within a few 100 Myr.  
This is because 
the average mass in our simulations is higher than it would be assuming a realistic mass spectrum, 
increasing the efficiency of gas damping.  We did attempt to perform some simulations adopting a 
realistic mass spectrum, and our preliminary results confirm these conclusions.

\subsection{Model assumptions} \label{assumptions}

In this section, we improve upon the connection between our results and real astrophysical 
environments by discussing our model assumptions.

%First, when do the initial conditions adopted in our model actually occur in nature?  In 
%particular, although the issue is still being debated, it has long been thought that the star 
%formation process should result 
%in primordially mass segregated clusters, with the most massive stars located in the core.  
%
%One example is after a phase of violent relaxation, as might occur when a cluster merges or collides 
%with a gas-rich galactic nucleus.  

%First, our model assumes that initially all mass species have the same root-mean-square speed.  This 
%assumption should be suitable to a cluster that has recently undergone a phase of violent relaxation.  In 
%open clusters, this could occur if the parent star-forming region is initially sub-virial, and a 
%pronounced infall phase occurs.  This could also apply to the formation of nuclear star clusters in 
%galactic nuclei.  However, here, violent relaxation could also occur when significant reservoirs of 
%gas fall into the nucleus, 
%or when/if other star clusters, in particular massive globular clusters, spiral in and merge 
%with the nucleus due to stellar dynamical friction.

First, we discuss our derivation for the timescale for two-body relaxation to 
operate in a gas-embedded star cluster, as presented in Equation~\ref{eqn:t-rh-gas}.  Importantly, 
we have modeled the gas as a simple background potential in this derivation, which neglects the 
gravitational interactions between stars and over-densities in the gas such as, for example, any 
filamentary structure in the gas.  
%With 
%that said, assuming the turbulent motions of the gas (presumably giving rise to these over-densities) 
%are directed randomly with respect to the 
%stellar orbits, then this could be a decent first-order approximation.  
We do not expect our model to be 
accurate in the regime where the total gas mass significantly exceeds the total stellar mass, since 
here the assumption that the cluster is in steady-state breaks down, as does the assumption that the gas 
can be treated as a simple background potential.  More detailed numerical 
simulations will be needed to identify the parameter space suitable to our simple analytic treatment for 
a gas-modified relaxation time.  

Next, we discuss our simplified treatment of gas dynamical friction and accretion.  Probably the most 
important 
assumption is that the radial density profile of the gas is uniform, which is certainly not the case 
in most observed star-forming regions.  Among other things, the gas could be more 
centrally concentrated than the stars due to dissipation, and/or it could exhibit over- and under-densities 
due to radiation pressure from stars and its own self-gravity.  If a 
region is of sufficiently low density, 
perturbations traveling through the gas may not be able to propagate through it, and this would 
decrease the effectiveness of gas dynamical friction.  What's more, 
the orbits of neighboring stars pass directly through the wakes induced by gas dynamical friction.  This 
suggests that the true upper limit for the Coulomb logarithm, incorporated in the derivation for the 
deceleration due to gas dynamical friction in the supersonic regime, should perhaps be the distance 
between stars, instead of the cluster half-mass radius.  This reduces the rate of deceleration due to 
gas dynamical friction by a factor $\lesssim 10$.  
%
%The assumption of a uniform gas density is also adopted in the derivation for the rate of deceleration 
%due to gas accretion.  In particular, 
These issues are also not accounted for by our simple estimate for the accretion rate.  
% assumption that the accretion stream should be symmetric about 
%the direction of motion of the accretor, as adopted in the Bondi-Hoyle-Lyttleton approximation, is likely also 
%not valid in real astrophysical situations.  This assumption affects the calculation of the amount of 
%matter that becomes bound to the accretor, by maximizing the amount of energy and angular momentum dissipated 
%by gravitationally-focused gas colliding along the axis of motion of the accretor.  The 
The Bondi-Hoyle-Lyttleton prescription for accretion represents a strict upper limit, and the true accretion rate 
could be orders of magnitude lower \citep[e.g.][]{krumholz04,krumholz05a,krumholz06}.  

Our model is only suited to the steady-state subsonic and supersonic limits, or $v \ll c_{\rm s}$ 
and $v \gg c_{\rm s}$, respectively.  Thus, the assumptions 
behind our analytic timescales break down for stellar velocity 
dispersions on the order of the sound speed.  Here, gas dynamical friction
is at its most efficient, and could be the dominant damping force acting on the
test particle for much lower gas densities than our results suggest.  

Given the simplicity of our derivations for the mass segregation times due to gas
accretion and gas dynamical friction, we estimate they are correct to within an order of 
magnitude, at best.  In fact, given the (flawed) assumption of a uniform gas density, the 
derived rates are likely over-estimates of the true rates.  %Since these time-scales differ by at most this 
%amount, 
In general, the conclusion that the gas dynamical friction timescale is shorter than the gas 
accretion timescale for all but the most bloated objects requires further study using more 
advanced simulations.  Indeed, \citet{lee11} and \citet{lee13} recently argued
that the deceleration due to accretion cannot be separated from that due to gas dynamical
friction, and that the damping force is precisely equal to $\dot{m}$v in both the subsonic and
supersonic regimes.

%Despite the simplicity of our analytic model, it gives 
%good agreement with the results of our computational $N$-body models.  With that 
%said, it is of course important to bear in mind that, by construction, both models are founded 
%on some of the same basic assumptions.  

To derive the timescales for gas dynamical friction and accretion to cause a massive test 
particle to become mass segregated, we assumed that both gas damping and two-body relaxation 
cause a particle's velocity to vary smoothly from an initial value $\sigma$ to a final 
value $\sigma\sqrt{\bar{m}/m_{\rm 1}}$.  This is not strictly valid, since particles do not 
adhere to circular orbits in clusters, so that their velocities will vary over the course 
of a crossing time.  In turn, the deceleration due to both gas dynamical friction and accretion 
should vary over a crossing time, particularly in the supersonic limit, since here both rates 
depend on the particle velocity.  This issue 
is addressed directly via our computational $N$-body models, which put the validity of this 
assumption to the test.  Given the reasonably good agreement (to within an order of magnitude) 
between our analytic model and the 
results of our simulations, we conclude that the assumption that a particle's velocity varies 
smoothly as it decelerates due to gas damping is a decent first-order approximation.

\subsection{Specific astrophysical environments} \label{specific}

When do the initial conditions adopted in our analytic model actually occur in nature?  In
particular, our model assumes that initially all mass species have the same root-mean-square speed.  This
assumption should be suitable to a cluster that has recently undergone a phase of violent relaxation.  In
open clusters, this could occur if the parent star-forming region is initially sub-virial, and a
pronounced infall phase occurs.  This could also apply to the formation of nuclear star clusters in
galactic nuclei.  However, here, violent relaxation could also occur when significant reservoirs of
gas fall into the nucleus, 
or when/if other star clusters, in particular massive globular clusters, spiral in and merge
with the nucleus due to stellar dynamical friction within the host galaxy.

In all galactic nuclei and young star-forming regions considered in this paper, the stellar 
motions should typically be in the supersonic regime, since either the gas is predominantly cold 
and molecular or the stellar velocity dispersion is very high.  Nevertheless, it is possible that 
in some environments, the gas is sufficiently 
heated by, for example, stellar winds, supernovae, or high-energy radiation, that the gas 
sound speed becomes on the order of the stellar velocity dispersion.  Indeed, the gas in the 
Galactic Centre within $\sim$ 1 pc of the SMBH is known to be hot (kT $\sim$ 1 keV) \citep{merritt13}, 
however here the stellar velocity dispersion is also high.  
We note that, when the stellar velocities are on the order of the sound speed, gas damping should 
be maximally effective, since the deceleration 
due to gas dynamical friction becomes very large \citep[e.g.][]{ostriker99}.  

Below, we discuss in more detail the implications of our results for specific 
astrophysical environments.

\subsubsection{Galactic nuclei and primordial globular clusters} \label{gal-nuc}

The results of our analytic model suggest that, although the rate of gas damping should 
increase with increasing gas density, the effect is largely canceled by the additional gas mass 
causing an increase in the stellar velocity dispersion.  This suggests that the total duration of the 
gas-embedded phase plays a more important role in deciding the overall effects of gas damping 
on the cluster structure than does the gas density.  In late-type galaxies, galactic nuclei are 
thought to undergo continual gas replenishment from the host galaxy.  
If a new stellar population is born every time this occurs, a correlation could exist between the 
number of distinct stellar populations and the overall compactness of the nuclear cluster, 
assuming that the duration of each gas-embedded phase stays roughly the same between gas 
replenishment events.  Additionally, there is some
evidence to suggest that more massive GCs are more 
likely to host multiple stellar populations \citep[e.g.][]{gratton12}.  This is in rough
agreement with the idea that more compact (or 
concentrated) clusters should also host a larger fraction of second or even third generation stars,
or perhaps just a wider stellar age distribution if distinct generations are not present.

As discussed, gas damping should cause clusters to contract, independent 
of the mass segregation process.  The effect can be significant.  For example, 
in our simulations with gas damping, we find a factor of $\sim 2$ difference in the final cluster 
half-mass radius compared to those simulations without gas damping, for our chosen model 
assumptions.

In galactic nuclei, cluster contraction could be relevant to SMBH 
formation in the early Universe, and even SMBH growth at later cosmic epochs.  If an SMBH is 
present at the centre of a nuclear star cluster, gas damping will increase the feeding rate of stars 
into its immediate vicinity, and hence the rate at which stars merge with it.  More generally, 
gas damping could accelerate a phase of runaway mergers to occur in the centres of 
nuclear clusters, 
which could then result in SMBH formation or growth.  The simulations performed in this paper treat all 
objects as point particles, so that mergers/collisions are neglected.  However, modern simulation-based 
techniques can address this issue \citep[e.g.][]{portegieszwart04}, which offers an interesting avenue 
for future studies.  Similarly, our results can be applied to stellar-mass black holes in primordial 
globular clusters, which suggests that gas damping could have an important bearing on the present-day 
black hole retention fractions.

\textbf{We caution that we have not considered stellar evolution-induced mass loss in our simulations, 
which contributes to cluster expansion.  This is particularly important 
early on in the cluster lifetime, when massive stars are still present \citep[e.g.][]{chernoff90,leigh13b}.  
We have also neglected dynamical interactions involving binary stars, which can act either as a 
source of heating or cooling \citep[e.g.][]{fregeau09,converse11,leigh13b}.  Future more sophisticated 
simulations should ideally account for these processes, and their implications for the results 
presented here.}   

%In primordial GCs, the phase of accelerated contraction could help push clusters to core collapse 
%on much shorter timescales than is done by two-body relaxation alone.  In general, if the duration 
%of the gas-embedded phase increases with increasing 
%total cluster mass, then this should contribute to a correlation between the cluster concentration 
%and the total cluster mass, as discussed in \citet{leigh13}.  This could help to explain why more 
%massive MW GCs tend to be more concentrated \citep{harris96}, which would have important implications 
%for the results of \citet{demarchi07}.  These authors 
%found a relation between the slope of the present-day mass function and the concentration 
%parameter that cannot be explained by dynamical evolution alone, assuming a universal initial 
%mass function.  This relation can be explained, however, if an initial correlation exists 
%between the total cluster mass and concentration \citep{leigh13}.  Additionally, there is some 
%evidence to suggest that more massive GCs are more 
%likely to host multiple stellar populations \citep[e.g.][]{gratton12}.  This is in rough 
%agreement with the idea presented in the previous paragraph, namely that more compact (or 
%concentrated) clusters should also host a larger fraction of second or even third generation stars, 
%or perhaps just a wider stellar age distribution if distinct generations are not present.

\subsubsection{Young open clusters} \label{open}

Most young open clusters are not yet in virial equilibrium, and hence are not in steady-state.  In 
some cases, the clusters are sub-virial, and are in the process of contracting.  This 
is exactly what we would expect if gas dynamical friction and/or accretion are acting.  Specifically, 
the stellar velocities should be lower than expected from the virial theorem given the currently 
observed cluster size and structure.  With gas dynamical friction and accretion operating, the stars 
should continually be decelerating, while also attempting to adjust the cluster structure accordingly.  
The former must take place before the latter can occur.  Thus, both gas dynamical friction and accretion 
could contribute to causing a gas-embedded cluster to appear sub-virial.

If star clusters are born primordially mass segregated, our analytic 
timescales do not strictly apply.  Indeed, several recent studies suggest that star-forming
regions could be mass segregated as early as the protostellar phase 
\citep[e.g.][]{kryukova12,elmegreen14,kirk14}.  However, 
with our results in mind, it is perhaps no surprise that many young open clusters appear 
primordially mass segregated, if this conclusion is based on the ages of the clusters being much shorter 
than their half-mass relaxation times.  All of the mechanisms contributing to mass segregation discussed 
in this paper should be operating \textit{during} the star formation process, and their rates could be 
relatively short compared to the star formation time.  As protostars accrete mass from the ISM, 
conservation of momentum should continually 
act to reduce (typically) the velocities of the protostars.  At the same time, protostars are being 
accelerated/decelerated by the gravitational tug from their peers.  The most massive protostars should 
experience the greatest overall deceleration due to momentum conservation from accretion, and/or 
experience the least overall acceleration from their less massive peers.  Thus, by 
the time a statistically significant distribution of protostar masses has formed, it 
stands to reason that a cluster could already appear mass segregated \citep[e.g.][]{girichidis12b}.  
This represents an 
extreme application of our model, and is better addressed using more sophisticated numerical 
simulations of star formation.  Previous SPH simulations have shown that stars in the 
central cluster regions tend to accrete the most due to the higher gas densities, and this is 
primarily responsible for star clusters appearing primordially mass segregated 
\citep{bonnell97,bonnell98,bonnell01}.  However, primordial mass segregation can be complicated if 
(massive) clusters are initially born with significant substructure and undergo one or more phases 
of clump-infall.  It is perhaps more 
likely that the initial conditions of our model would at some point be met if such a scenario were 
to occur \citep[e.g.][]{maschberger10}.

If our simulations 
were to be repeated assuming a more realistic gas density profile that follows that of the stars,
many of the effects discussed in this paper would be amplified, such as the acceleration of core collapse.  
In general, we do not consider the evolving properties of the gas, and could be missing a lot of 
interesting physics due to this simplifying assumption.  Our analytic model is meant to provide a
useful benchmark for comparison between observational data and future more sophisticated
numerical simulations.

\section{Summary} \label{summary}

In this paper, we study the effects of gas damping on the evolution of embedded star clusters.  
Using a simple three-component analytic model, we compare the rates of mass segregation 
due to two-body relaxation, accretion from the interstellar medium, and gas dynamical 
friction in both the supersonic and subsonic regimes.  
Using observational data in the literature, we apply our analytic predictions to two 
different astrophysical environments, namely galactic nuclei and young open star clusters.  The 
general predictions of our analytic model are confirmed using numerical $N$-body simulations, 
modified to include the effects of gas damping. 

The effects of gas damping can be significant in some gas-rich 
nuclei and even some gas-embedded low-mass open clusters, significantly reducing 
the timescale for mass segregation below that due to two-body relaxation alone.  However, two-body 
relaxation dominates in both environments considered here.  In general, 
our results suggest that gas damping is relatively inefficient in very massive clusters with long 
relaxation times, even when the gas density is high.  This is because the higher gas density
translates into a larger cluster mass, and thus larger
stellar velocities via the virial theorem.  

Gas damping causes overall cluster contraction.  If the cluster is mass segregated, the 
core radius contracts faster than the half-mass 
radius, increasing the central concentration and accelerating the rate of core collapse.  This effect 
should be further amplified if the gas density follows the stellar density, and is higher in the central 
cluster regions.  A stable state of approximate energy equilibrium cannot be maintained if gas 
damping is present, even if Spitzer's Criterion is satisfied.  This instability drives the continued dynamical 
decoupling and subsequent ejection (and/or collisions) of the more massive population.  

%In this paper, we do many things, none of which are particularly worth summarizing here.  
%Instead, we recant the lyrics from Cyndi Lauper's famous song entitled ``Girls Just 
%Wanna Have Fun!''...

\section*{Acknowledgments}

We kindly thank Eric Rosolowsky and Natasha Ivanova for useful discussions, as well as an 
anonymous referee whose comments helped to significantly improve our manuscript.

%\chapterbib                                                                                                                                            

\appendix \label{app}

\section{Analytic model} \label{analytic}

Consider a three-component model for a spherical gas-embedded self-gravitating system of 
massive particles.  The individual masses of 
the components are m$_{\rm 1}$, m$_{\rm 2}$ and m$_{\rm 3}$, and satisfy the relations 
m$_{\rm 1} >$ m$_{\rm 2} \gg$ m$_{\rm 3}$ and M$_{\rm 1} \ll$ M$_{\rm 2} \approx$ M$_{\rm 3}$.  
Here we let M$_{\rm 1}$, M$_{\rm 2}$ and M$_{\rm 3}$ represent the total masses in 
components 1, 2 and 3, respectively.  Spitzer's Criterion \citep{spitzer69} is satisfied 
for all permutations of species 1, 2 and 3.  Thus, although not guaranteed,\footnote{Whether 
or not a self-gravitating system of stars actually achieves energy equipartition is a complicated 
technical issue (see \citet{trenti13} for more details).  Spitzer's Criterion merely 
provides an approximate guide.} energy equipartition, or more accurately energy equilibrium, is 
in principle possible for our three-component system.  As discussed in Section~\ref{analytic}, energy
equipartition typically does not occur in real star clusters and a stable state of
energy \textit{equilibrium} tends to be achieved instead \citep[e.g.][]{trenti13}.  However, as 
we will show, the 
precise final velocity does not significantly influence our calculations for the different mass
segregation timescales.  Thus, we assume a final state of approximate energy equipartition
throughout our calculations for simplicity.  

Putting our model within the context of a real star 
cluster, components 1 and 2 constitute the total stellar mass M$_{\rm s} =$ M$_{\rm 1} +$ M$_{\rm 2}$, 
and component 3 constitutes the total gas mass M$_{\rm g} =$ M$_{\rm 3}$.  Instead of the average 
stellar mass, we use the \textit{root-mean-square} mass in species 1 and 2, denoted by $\bar{m}$ \citep{perets07}.  
The root-mean-square mass is given by:
\begin{equation}
\label{eqn:avg-s}
\bar{m} = \sqrt{\frac{N_{\rm 1}m_{\rm 1}^2 + N_{\rm 2}m_{\rm 2}^2}{N_{\rm 1} + N_{\rm 2}}}
%m_{\rm 1}m_{\rm 2}\frac{M_{\rm 1}+M_{\rm 2}}{m_{\rm 2}M_{\rm 1}+m_{\rm 1}M_{\rm 2}}\\
%        = \frac{M_{\rm 1}+M_{\rm 2}}{N_{\rm 1}+N_{\rm 2}},
\end{equation}
where N$_{\rm 1} =$ M$_{\rm 1}$/m$_{\rm 1}$ and N$_{\rm 2} =$ M$_{\rm 2}$/m$_{\rm 2}$ 
denote the number of objects belonging to species 1 and 2, respectively.
%We assume 
%that the gas is isothermal, distributed with a constant density profile, and composed 
%entirely of hydrogen (either ionized or molecular; see below).  

Now, consider a test particle of mass m$_{\rm 1}$ orbiting within the system.  
Initially, the velocity of the test particle v$_{\rm 1}$ is set equal to the root-mean-square 
speed of the system \citep{spitzer69,binney87}:
\begin{equation}
\label{eqn:vel-disp}
\sigma = \sqrt{\frac{2G(M_{\rm s}+M_{\rm g})}{5r_{\rm h}}},
\end{equation}
where r$_{\rm h}$ is the half-mass radius of the cluster.  In the subsequent sections, we will 
calculate the timescales for all three damping mechanisms to reduce the speed of the 
test particle from $\sigma$ to $\sigma\sqrt{\bar{m}/m_{\rm 1}}$, which signifies 
energy equipartition and, as discussed in the subsequent section, mass segregation.

\subsection{Two-body relaxation in gas} \label{2body-gas}

In this section, we calculate the rate for gravitational interactions between 
stars to bring a massive test particle orbiting within a gas-embedded system to an 
orbit that is roughly consistent with energy equilibrium.  We refer to 
this final state as ``mass segregated''.  We ignore short-range interactions between 
stars, such as energetic scattering events and direct collisions, 
since these do not become important until \textit{after} mass segregation has occurred, 
and the heaviest particles are confined to the bottom of the total cluster potential.  
%in systems for which the presence 
%of the gas plays a significant gravitational role.  
We focus only on long-range interactions, 
specifically modifying the rate of two-body relaxation in a gaseous medium, modeled as a 
background potential.\footnote{The assumption that the gas can be modeled as a simple 
background potential is likely not valid over the entire range of cluster masses considered 
here.  To first order, we expect it to breakdown when the total stellar mass is much smaller 
than the total gas mass, or M$_{\rm s} \ll$ M$_{\rm g}$.  We will return to this issue in 
Section~\ref{discussion}.}   Here, 
the terms short- and long-range refer to distances on the order of the stellar and cluster 
(i.e. half-mass) radii, respectively.

%WHAT ABOUT STELLAR DYNAMICAL FRICTION, WHICH IS MORE A SHORT-RANGE PHENOMENON INVOLVING 
%STAR-STAR INTERACTIONS?  OR IS THIS IRRELEVANT, SINCE TWO-BODY RELAXATION, AND THUS LONG-RANGE 
%ENCOUNTERS, DOMINATE OVER SHORT-RANGE ENCOUNTERS?
%WHAT IS THE DIFFERENCE BETWEEN STELLAR DYNAMICAL FRICTION AND TWO-BODY RELAXATION IN A CLUSTER 
%WITH A REAL MASS SPECTRUM?  ARE THEY NOT THE SAME THING, ALTHOUGH THEY WOULDN'T BE IF TWO-BODY 
%RELAXATION WERE OCCURRING IN A SINGLE-MASS CLUSTER, SINCE HERE STELLAR DF WOULD NOT OCCUR?
%
%NOTE:  THE STELLAR DYNAMICAL FRICTION TIME IS EQUIVALENT TO THE COLISIONLESS GAS DYNAMICAL 
%FRICTION TIME, WHICH HAS A MINIMUM NEAR V ~ C_S.  OUR INITIAL CONDITIONS ARE SUCH THAT THE 
%MASSIVE TEST PARTICLE HAS A VELOCITY NEAR THE VELOCITY DISPERSION OF THE SYSTEM, WHICH IS 
%EQUIVALENT TO V ~ C_S.  THUS, THE STELLAR DYNAMICAL FRICTION TIME MIGHT ACTUALLY BE SHORTER 
%THAN THE TWO-BODY RELAXATION TIME FOR OUR CHOSEN INITIAL CONDITIONS.  MUST CALCULATE IT!!!
%
%\subsubsection{Two-body relaxation in gas} \label{2body}

We wish to calculate the time required for the test particle to be decelerated 
to a mean speed $\sqrt{\bar{m}/m}\sigma$ due to star-star gravitational interactions only.  In 
the absence of gas, this time 
corresponds to the equipartition time, which is roughly equal to the 
relaxation time \citep[e.g.][]{heggie03}.  For the test particle, this time 
is given by \citep{vishniac78}:
\begin{equation}
\label{eqn:tau-relax}
\tau_{\rm rh}(m_{\rm 1}) = \frac{\bar{m}}{m_{\rm 1}}\tau_{\rm rh},
\end{equation}
where
\begin{equation}
\label{eqn:t-rh}
\tau_{\rm rh} [yr] = 1.7 \times 10^5 N_{\rm s}^{1/2}\Big( \frac{r_{\rm h}}{1 {\rm pc}} \Big)^{3/2} \Big( \frac{1 M_{\odot}}{\bar{m}} \Big)^{1/2},
\end{equation}
and N$_{\rm s} =$ N$_{\rm 1} +$ N$_{\rm 2}$ is the total number of stars.

%Call the below a theorem?  Proof:  Poisson's equation remains unchanged, 
%since the constant drops out.
Is this approximation still valid in the presence of gas?  We will argue that, 
if the \textit{shape} of the total gravitational potential is approximately 
the same with or without the gas, then the rate at which two-body relaxation 
operates on the stellar mass is the same to within a dimensionless constant, which is 
equal to the ratio between the total mass in gas and stars.  In this approximation, 
the gas is treated as a simple background potential.

Consider a gravitational potential $\Phi_{\rm s}$ that describes 
% (or observable two-dimensional density 
%profiles) from which can be derived distribution functions that are 
an unrelaxed, self gravitating, steady-state (i.e. virialized) stellar system, based on the 
corresponding 
solution to Jeans' equation (see \citet{lynden-bell62b} for examples of such potentials).  
%,lynden-bell62b,lynden-bell62c}.  Provided such a potential accurately describes the 
If we include a gas component with potential 
%$\Phi_{\rm g}$, our total potential becomes:
%\begin{equation}
%\label{eqn:pot-tot}
%\Phi_{\rm tot} = \Phi_{\rm s} + \Phi_{\rm g}
%\end{equation}
%Assuming that the stellar and gas potentials are proportional, 
$\Phi_{\rm g}$ $= {\alpha}\Phi_{\rm s}$ where $\alpha > 0$, then 
$\Phi_{\rm g} + \Phi_{\rm s}$ is also a solution to Jeans' equation, and describes a 
steady-state, unrelaxed system.\footnote{This is only valid in the 
supersonic regime, where pressure forces are negligible in the gas and the collisionless 
Boltzmann equation applies.  In the subsonic regime, we simply assume that steady-state is achieved.}  
%Thus, $\Phi_{\rm tot}$ also describes 
%a steady-state, unrelaxed system.  In this case, the gas serves only to increase the stellar velocities 
%by increasing the depth of the potential (via the virial theorem).  
%Importantly, \textit{this does not 
%influence the rate of two-body relaxation}.  This is evident from ... PROVE IT!!!

%The additional gas mass translates into an increase in the relaxation time of the stellar 
%system by a factor (1+$\alpha$)$^{1/2}$.  This can be understood as follows.  From Poisson's equation, the 
%density profiles of the gas and stars are related by the constant $\alpha$, 
%i.e. $\rho_{\rm g} = {\alpha}\rho_{\rm s}$.
From Poisson's equation, we have M$_{\rm g} =$ ${\alpha}$M$_{\rm s}$ for the relation between the 
total mass in gas and stars.\footnote{Note that we do not specify a functional form 
for the radial density profile, and will work with the average cluster gas density from here on out.}  
Thus, with these assumptions, we can replace the total mass in stars M$_{\rm s} =$ N$\bar{m}$ in 
Equation~\ref{eqn:t-rh} with (1+$\alpha$)M$_{\rm s}$.  This gives:
%Integrating over the spatial extent of the cluster gives M$_{\rm g} =$ ${\alpha}$M$_{\rm s}$ for 
%the relation between the total mass in gas and stars.  
%Substituting M$_{\rm tot} =$ M$_{\rm g} +$ M$_{\rm s} =$ (1+$\alpha$)M$_{\rm s}$ for the total 
%stellar mass M$_{\rm s} =$ N$\bar{m}$ in Equation 4-7 of \citet{binney87} and using the 
%virial theorem to remove the dependence on velocity, we obtain an 
%additional factor (1+$\alpha$)$^{1/2}$ in Equation 4-9, which provides a simple expression for 
%the relaxation time.  
%Using the virial theorem to remove the dependence on the particle 
%velocity 
%
%Thus, in the presence of gas, Equation~\ref{eqn:t-rh} must be modified to:
\begin{equation}
\label{eqn:t-rh-gas}
\tau_{\rm rh} [yr] = 1.7 \times 10^5 (1 + \alpha)^{1/2}N_{\rm s}^{1/2}\Big( \frac{r_{\rm h}}{1 {\rm pc}} \Big)^{3/2} \Big( \frac{1 M_{\odot}}{\bar{m}} \Big)^{1/2},
\end{equation}
where $\alpha$ is the ratio between the total mass in gas and stars, or $\alpha =$ M$_{\rm g}$/M$_{\rm s}$.  
This form for the half-mass 
relaxation time should be reasonable late in the star formation process, after a significant 
fraction of the gas mass has been converted to stars.

%What if $\Phi_{\rm g} \ne {\alpha}\Phi_{\rm s}$?  This question can be answered within the 
%context of our previous assumptions by 
%asking:  What happens if one of the potentials is perturbed?  The answer can be found using 
%perturbation theory by 
%solving Jeans' equation for both potentials, as well as their sum, simultaneously.  

\subsection{Accretion from the ISM} \label{accretion}

In this section, we consider how predominantly short-range (i.e. comparable to the stellar radius) 
gravitational interactions between stars and gas 
contribute to accelerating the rate of mass segregation above the effects 
of two-body relaxation (i.e. long-range star-star interactions) alone.  To this end, we consider two 
additional damping mechanisms 
that could operate on a test particle of mass m.\footnote{Throughout this section, we replace m$_{\rm 1}$ 
by the more general variable m, since the derived timescales apply to any massive celestial object.}  
These are gas dynamical friction and accretion from the ISM.  

We consider gas dynamical friction and gas accretion as being entirely independent mechanisms.  
The drag induced by accretion corresponds 
to the momentum imparted to the accretor from the accreting gas, so that 
the gas is co-moving with the accretor and gravitationally bound to it.  The drag induced by gas 
dynamical friction, on the other hand, arises due to an asymmetry in the 
gas flow along the axis of motion of the perturber.  Typically, a wake is formed 
downstream of the perturbing object which exerts a gravitational tug on it, thereby 
reducing its speed.  As we will show, in each of the subsonic 
and supersonic regimes, the derived timescales for accretion and gas dynamical friction 
are equivalent to within a dimensionless constant, for a given test particle mass.  This 
is because both timescales are derived assuming spherically symmetric perturbations on the 
gas, with the 
degree of damping arising due to the asymmetry induced by the motion of the perturbing 
object relative to the gas.  With that said, we note that \citet{lee11} recently showed that the 
rate of momentum damping due to gas dynamical friction is precisely equal to $\dot{m}$v, and 
that accretion and gas dynamical friction cannot be separated with their formalism into separate 
damping mechanisms.  Regardless, the derived 
estimates for the rate of deceleration of a particle due to gas dynamical friction by \citet{lee13} 
agree well with those of \citet{ostriker99} and previous authors in the steady-state supersonic 
limit.
%The relevant asymmetry is located upstream for accretion, 
%and downstream for gas dynamical friction.

%\textbf{Are the derivations in the subsequent section only applicable to the adiabatic case?  - i.e. the 
%perturbing mass does not have enough time to re-adjust its velocity to match the orbital 
%velocity at a given distance from the cluster centre?  Otherwise, we need to integrate over the orbit
%of the object, in order to know how its velocity changes in response to damping, which in turn 
%affects the relevant time-scales.  Instead, we assume that the damping mechanisms operate on a 
%time-scale that is comparable to the orbital period...
%I think we need an assumption like this to justify integrating from v to $\sqrt{\bar{m}/m}v$, and 
%hence this being an adequate assumption...  Maybe it is okay to simply to discuss this in the 
%Discuss section by describing how the $N$-body models we run cover this weak assumption, and show that 
%it actually gives a reasonable approximation despite its simplicity?}
%
%\subsubsection{Accretion from the ISM} \label{accretion}
\subsubsection{The subsonic limit} \label{subsonic1}

The time required for accretion to decelerate the particle 
by roughly the same amount as is done by two-body relaxation in a single relaxation 
time was derived in \citet{leigh13a} for the subsonic limit.  For an accretion rate:
\begin{equation}
\label{eqn:acc-rate}
\dot{m} = {\lambda}{\delta}m^{\epsilon},
\end{equation}
this timescale is given by:
\begin{equation}
\label{eqn:tau-acc}
\tau_{\rm acc} = \frac{m^{1-\epsilon}\Big( (m/\bar{m})^{(1-\epsilon)/2} - 1 \Big)}{{\delta}(1 - \epsilon)},
\end{equation}
where $\epsilon > 1$, and $\lambda$ and $\delta$ are coefficients that determine the accretion rate.  We 
set $\lambda = 1.0$ for the remainder of this paper, and absorb this factor into the coefficient $\delta$.  
For Bondi-Hoyle accretion, $\epsilon=2$ and we then have for the coefficient \citep[e.g.][]{bondi44,maccarone12}:
\begin{equation}
\label{eqn:delta}
\delta = 7 \times 10^{-8} {M_{\odot}}^{-1}{\rm yr}^{-1} \Big( \frac{n}{\rm 10^6 cm^{-3}} \Big) \Big( \frac{\sqrt{c_{\rm s}^2 + v^2}}{\rm 10^6 cm s^{-1}} \Big)^{-3},
\end{equation}
where $n$ is the particle number density, $c_{\rm s}$ is the gas sound speed, and $v$ is the velocity 
of the accretor relative to the gas.  Plugging $\epsilon=2$ into Equation~\ref{eqn:tau-acc} gives:
\begin{equation}
\label{eqn:tau-acc1}
\tau_{\rm acc} [yr] = 1.4 \times 10^8 \Big( 1 - \frac{\bar{m}}{m} \Big)^{-1/2} \Big( \frac{10^6 cm^{-3}}{n} \Big) \Big( \frac{\sqrt{c_{\rm s}^2 + v^2}}{\rm 10^6 cm s^{-1}} \Big)^{3} \Big( \frac{1 M_{\odot}}{m} \Big).
\end{equation}

Importantly, Equation~\ref{eqn:tau-acc1} is only valid in the subsonic regime when 
$v \lesssim c_{\rm s}$, since it assumes that the velocity of the accretor relative to the gas 
remains roughly constant as 
it accretes.  Thus, it corresponds to an upper limit, since accretion serves to reduce the accretor's 
speed, which 
in turn accelerates the rate of accretion.  
%This assumption, and hence Equation~\ref{eqn:tau-acc2}, is 
%only valid provided $v \lesssim c_{\rm s}$.  
In the limit $v \ll c_{\rm s}$, we can modify Equation~\ref{eqn:tau-acc1} by dropping the 
velocity term:
\begin{equation}
\label{eqn:tau-acc2}
\tau_{\rm acc,sub} [yr] = 1.4 \times 10^8 \Big( 1 - \frac{\bar{m}}{m} \Big)^{-1/2} \Big( \frac{10^6 cm^{-3}}{n} \Big) \Big( \frac{c_{\rm s}}{\rm 10^6 cm s^{-1}} \Big)^{3} \Big( \frac{1 M_{\odot}}{m} \Big).
\end{equation}

%Averaging over a Maxwellian velocity distribution to obtain a time-scale that is appropriate to 
%the entire population of species 1 (i.e. the heavier species), we obtain:

\subsubsection{The supersonic limit} \label{supersonic1}

If the motion is in the supersonic 
regime, i.e. $v \gtrsim c_{\rm s}$, then changes in the accretor velocity can have an important bearing 
on the accretion rate.  Hence, in this case, we must also integrate over the relative velocity between
the gas and the accretor.  In \citet{leigh13a}, this was not necessary, since we considered only a single 
(high) value for the sound speed, and the motion was not supersonic.  In 
this paper, however, we are interested in a range of sound speeds, and must deal with supersonic 
motion.  Thus, we must re-derive Equation~\ref{eqn:tau-acc1} to obtain an analogous expression in the 
supersonic regime for the approximate time required for the test particle to become mass segregated.
We assume $v \gg c_{\rm s}$ throughout this derivation in order to ensure a consistent comparison 
to the corresponding timescale for gas dynamical friction, derived in the subsequent 
section.\footnote{We do not address the case $v \sim c_{\rm s}$ analytically, since the assumptions 
adopted for the subsonic and supersonic regimes do not apply here, and the complexity of the 
problem increases greatly.}

%Writing the accretor velocity as a function of time using conservation of momentum, we have:
%\begin{equation}
%\label{eqn:vel-time}
%v(t) = \frac{m}{m(t)}v.
%%%%%\int_{0}^{\tau_{\rm acc}} \dot{m}(t)dt = m_{\rm f} - m,
%\end{equation}
%We can take $\delta \propto v^{-3}$ in Equation~\ref{eqn:delta} for the supersonic regime, since 
%$v \gg c_{\rm s}$.  Hence, 
We write:
%$m_{\rm f}$ is the final mass of the accretor.  
\begin{equation}
\label{eqn:tau-acc3}
\tau_{acc} = \int_{m}^{\sqrt{m^3/\bar{m}}} \frac{dm}{\dot{m}}.
\end{equation}
We can take $\delta \propto v^{-3}$ in Equation~\ref{eqn:delta} for the supersonic regime, since
we assume $v \gg c_{\rm s}$.  Hence, plugging Equations~\ref{eqn:delta} and~\ref{eqn:acc-rate} into 
Equation~\ref{eqn:tau-acc3} and using conservation of momentum, we obtain for the time needed 
for accretion to reduce the test 
particle's speed from $\sigma$ to $\sqrt{m/\bar{m}}\sigma$ in the supersonic regime:
\begin{equation}
\label{eqn:tau-acc4}
\tau_{\rm acc,sup} [yr] = 3.5 \times 10^7 \Big( 1 - \Big( \frac{\bar{m}}{m} \Big)^2 \Big) \Big( \frac{\rm 10^6 cm^{-3}}{n} \Big) \Big( \frac{v}{\rm 10^6 cm s^{-1}} \Big)^{3} \Big( \frac{1 M_{\odot}}{m} \Big).
\end{equation}

\subsubsection{Direct accretion} \label{direct}

In the limit where the test particle velocity exceeds the escape speed from its surface, 
Bondi-Hoyle-Lyttleton accretion no longer applies.  This is because 
the radius of the object becomes larger than the gravitationally-focused cross-section.  
Thus, in this regime, it is 
the direct plowing of the gas by the test particle that reduces its momentum, and 
hence speed.  We refer to this as \textit{direct accretion}.  This scenario is the most relevant 
to low-mass, bloated objects orbiting 
in an environment with a high velocity dispersion.  The timescale required for direct 
accretion to bring a test particle into approximate energy equipartition 
%is:
%\begin{equation}
%\label{eqn:tau-acc5}
%\tau_{\rm acc,dir} [yr] = 1.2 \times 10^{15} \ln\Big( \frac{m}{\bar{m}} \Big) \Big( \frac{\rm 10^6 cm^{-3}}{n} \Big) \Big( \frac{\rm 10^6 cm s^{-1}}{v} \Big) \Big( \frac{1 R_{\rm \odot}}{R} \Big)^2 \Big( \frac{m}{1 M_{\odot}} \Big),
%\end{equation}
%where $R$ is the physical radius of the test particle or accretor.  To derive Equation~\ref{eqn:tau-acc5}, 
can be calculated as the time required for an object of cross-sectional area ${\pi}R^2$ to collide with sufficient gas 
mass to reduce its speed from $\sigma$ to $\sqrt{m/\bar{m}}\sigma$ using conservation of momentum.  
Using this approximation, we calculate a timescale on the order of 100 Myr for an object with m $\sim$ 1 M$_{\rm \odot}$ 
and R $\sim$ 1000 R$_{\rm \odot}$, which is applicable to a star in the asymptotic giant branch (AGB) phase of evolution.  
This is shorter than the corresponding timescale for gas dynamical friction, but 
longer than the typical duration of the AGB phase.  Assuming instead R $\sim$ 100 R$_{\rm \odot}$, this timescale is on the 
order of a Hubble time.  Thus, we do not expect direct accretion to dominate the deceleration of any 
test particle over the majority of its lifetime.  Importantly, our derivation ignores AGB winds and radiation, and how these 
interact with the surrounding ISM prior to direct accretion, which could increase $\tau_{\rm acc,dir}$ 
substantially.  Thus, for these reasons, we do not concern ourselves with the direct accretion scenario in the 
subsequent sections.  
%However, we will return to it briefly in Section~\ref{discussion}.

\subsection{Gas dynamical friction} \label{gas-dyn}

To derive the time required for gas dynamical friction to decelerate the test particle 
from $\sigma$ to $\sqrt{\bar{m}/m}\sigma$, we use the gas dynamical friction force taken 
from \citet{ostriker99}:
\begin{equation}
\label{eqn:fric-force}
F_{\rm df} = ma_{\rm df} = -F_{\rm 0}I,
\end{equation}
where $I$ is a function that depends on the ratio between the perturber velocity and the 
gas sound speed $c_{\rm s}$, which changes depending on whether the motion is subsonic or 
supersonic, and: 
\begin{equation}
\label{eqn:fric-coeff}
F_{\rm 0} = \frac{4{\pi}(Gm)^2\rho_{\rm g}}{v^2},
\end{equation}
where $v$ is again the velocity of the perturber relative to the gas, and $\rho_{\rm g}$ 
is the gas density.  

Using Equation~\ref{eqn:fric-force} for the deceleration, we have for the total time:
\begin{equation}
\label{eqn:tau-df1}
\tau_{\rm df} = \int_{v_{\rm i}}^{v_{\rm f}} \frac{dv}{a_{\rm df}}.
\end{equation}
where $v_{\rm i} = \sigma$ and $v_{\rm f} = \sqrt{\bar{m}/m}\sigma$.  We must 
integrate over the test particle velocity since the deceleration induced 
by the gas dynamical friction force is a function of $v$.  We note that 
we assume that the particle mass remains constant throughout this calculation, since we 
are treating accretion and gas dynamical friction independently.  

%This is because we are 
%interested in solving for the parameter space for which each of the different 
%mechanisms for decelerating the test particle dominates.  Thus, we are only 
%concerned with gas dynamical friction in the regime that accretion has a 
%negligible impact on the test particle velocity.  
%%%For low cluster masses, this 
%%%occurs for small sound speeds and comparably small relative velocities between 
%%%the accretor and the gas.  For large cluster masses, this occurs for 

The function $I$ depends on the velocity of the test particle relative to 
the sound speed, called the Mach number, and takes on two different forms depending 
on whether the motion 
is subsonic or supersonic.  In the subsonic regime, i.e. $v < c_{\rm s}$, we have 
\citep{ostriker99}:
\begin{equation}
\label{eqn:I-sub}
I_{\rm sub} = \frac{1}{2}\ln{\frac{c_{\rm s}+v}{c_{\rm s}-v}}-v/c_{\rm s}.
\end{equation}
In the supersonic regime, i.e. $v > c_{\rm s}$, we have:
\begin{equation}
\label{eqn:I-sup}
I_{\rm sup} = \frac{1}{2}\ln{\Big(1 - \frac{c_{\rm s}^2}{v^2}\Big)} + \ln{\Big(\frac{r_{\rm max}}{r_{\rm min}}\Big)},
\end{equation}  
where the last term corresponds to the Coulomb logarithm.  We take $r_{\rm max} = r_{\rm h}$ and 
$r_{\rm min} = R$, where $R$ is the physical radius of the test particle.  Note that for large 
radii $R$, the Coulomb logarithm is smaller and so is the corresponding gas dynamical friction timescale.

\subsubsection{The subsonic limit} \label{subsonic2}

First, we solve for the time needed for the speed of the test particle to be 
reduced to $\sqrt{\bar{m}/m}\sigma$ in the subsonic limit.  To do this for any velocity, 
we must plug Equation~\ref{eqn:I-sub} into Equation~\ref{eqn:tau-df1} and integrate
with respect to $v$.  The integration must be performed numerically if $v \lesssim c_{\rm s}$, 
and provides the desired timescale for gas dynamical friction in the subsonic regime, 
%$\tau_{\rm df,sub}$, 
as a function of the gas density and root-mean-square speed of 
the stellar system.  If gas dynamical friction is highly subsonic, i.e. $v \ll c_{\rm s}$, then 
%Don't we want to be discussing when it is efficient?  Maybe accretion happens sufficiently 
%fast we can make some simplifying assumption...? 
$I_{\rm sub}$ approaches $(v/c_{\rm s})^3/3$, and the drag force is proportional to the 
perturber's velocity.  Thus, in the limit of a very slow perturber, a simple analytic timescale 
for gas dynamical friction can be derived:
\begin{equation}
\label{eqn:tau-df2}
\tau_{\rm df,sub} [yr] = 2.6 \times 10^8 \Big(\ln\Big( \frac{m}{\bar{m}} \Big)\Big)^{-1} \Big( \frac{10^6 cm^{-3}}{n} \Big) \Big( \frac{c_{\rm s}}{10^6 cm/s} \Big)^3 \Big( \frac{1 M_{\odot}}{m} \Big).
\end{equation}
%Thus, this time-scale corresponds to
%an upper limit for the effectiveness of gas dynamical friction in the subsonic regime.

\subsubsection{The supersonic limit} \label{supersonic2}

To obtain the timescale for gas dynamical friction in the 
supersonic limit, or $\tau_{\rm df,sup}$, the procedure is exactly analogous as in 
the subsonic regime, except that it is Equation~\ref{eqn:I-sup} that is plugged into 
Equation~\ref{eqn:tau-df1}.  Again, the integration must be performed numerically if 
$v \gtrsim c_{\rm s}$.  However, 
as the motion becomes highly supersonic, i.e. $v \gg c_{\rm s}$, then $I_{\rm sup}$ approaches 
$\ln{r_{\rm max}/r_{\rm min}}$ and the drag force is proportional to $v^{-2}$.  Thus, 
in the limit of a very fast perturber, the timescale for gas dynamical friction becomes:
\begin{equation}
\label{eqn:tau-df3}
\tau_{\rm df,sup} [yr] = 2.9 \times 10^6 \Big( 1 - \Big(\frac{\bar{m}}{m} \Big)^{3/2} \Big) \Big( \frac{10^6 cm^{-3}}{n} \Big) \Big( \frac{v}{10^6 cm/s} \Big)^3 \Big( \frac{1 M_{\odot}}{m} \Big),
\end{equation}
where here we have used $\ln{r_{\rm max}/r_{\rm min}} = 10$ for the Coulomb logarithm.

\bsp

\label{lastpage}


\begin{thebibliography}{99}

%\bibitem[\protect\citeauthoryear{Baumgardt \& Makino}{2003}]{baumgardt03}
%Baumgardt H., Makino J. 2003, MNRAS, 340, 227
%\bibitem[\protect\citeauthoryear{Baumgardt, De Marchi \&
%    Kroupa}{2008}]{baumgardt08} Baumgardt H., De Marchi G., Kroupa
%  P. 2008, ApJ, 685, 247
\bibitem[\protect\citeauthoryear{Allison et al.}{2009a}]{allison09a} Allison R. J., Goodwin S. P., 
Parker R. J., Portegies Zwart S. F., de Grijs R., Kouwenhoven M. B. N. 2009, MNRAS, 395, 1449
\bibitem[\protect\citeauthoryear{Allison et al.}{2009b}]{allison09b} Allison R. J., Goodwin S. P.,
  Parker R. J., Portegies Zwart S. F., de Grijs R., Kouwenhoven M. B. N. 2009, ApJ, 700, 99
\bibitem[\protect\citeauthoryear{Bate, Bonnell \& Bromm}{2003}]{bate03} Bate M. R., Bonnell I. A., 
Bromm W. 2003, MNRAS, 341, 213
\bibitem[\protect\citeauthoryear{Bate}{2009}]{bate09} Bate M. R. 2009, MNRAS, 392, 590
\bibitem[\protect\citeauthoryear{Bate}{2012}]{bate12} Bate M. R. 2012, MNRAS, 419, 3115
\bibitem[\protect\citeauthoryear{Binney \& Tremaine}{1987}]{binney87}
  Binney J., Tremaine S., 1987, Galactic Dynamics (Princeton:
  Princeton University Press)
\bibitem[\protect\citeauthoryear{B\"oker, Lisenfeld \& Schinnerer}{2003}]{boker03} B\"oker T., 
Lisenfeld U., Schinnerer E. 2003, A\&A, 406, 87
\bibitem[\protect\citeauthoryear{Bondi \& Hoyle}{1944}]{bondi44} Bondi H.,
Hoyle F. 1944, MNRAS, 104, 273
\bibitem[\protect\citeauthoryear{Bondi}{1952}]{bondi52} Bondi H. 1952, MNRAS, 112, 195
\bibitem[\protect\citeauthoryear{Capuzzo-Dolcetta, Mastrobuono-Battisti \& Maschietti}{2011}]{capuzzo-dolcetta11} 
Capuzzo-Dolcetta R., Mastrobuono-Battisti A., Maschietti D. 2011, New Astronomy, 16, 284 
\bibitem[\protect\citeauthoryear{Bonnell et al.}{1997}]{bonnell97} Bonnell I. A., 
Bate M. R., Clarke C. J., Pringle J. E. 1997, MNRAS, 285, 201
\bibitem[\protect\citeauthoryear{Bonnell \& Davies}{1998}]{bonnell98} Bonnell I. A., 
Davies M. B. 1998, MNRAS, 295, 691
\bibitem[\protect\citeauthoryear{Bonnell et al.}{2001}]{bonnell01} Bonnell I. A.,
  Bate M. R., Clarke C. J., Pringle J. E. 2001, MNRAS, 323, 785
\bibitem[\protect\citeauthoryear{Chernoff \& Weinberg}{1990}]{chernoff90} Chernoff D. F.,
Weinberg M. D. 1990, ApJ, 351, 121
\bibitem[\protect\citeauthoryear{Conroy \& Spergel}{2011}]{conroy11} Conroy C., Spergel D. N.
2011, ApJ, 726, 36
\bibitem[\protect\citeauthoryear{Conroy}{2012}]{conroy12} Conroy C. 2012, ApJ, 758, 21
\bibitem[\protect\citeauthoryear{Converse \& Stahler}{2011}]{converse11} Converse J. M., 
Stahler S. W. 2011, MNRAS, 410, 2787 
\bibitem[\protect\citeauthoryear{Davies, Miller \& Bellovary}{2011}]{davies11} Davies M. B., 
Miller M. C., Bellovary J. M. 2011, ApJ, 740, 42
\bibitem[\protect\citeauthoryear{Dokuchaev}{1964}]{dokuchaev64} Dokuchaev V. P. 1964, 
Sovet Astron., 8, 23
\bibitem[\protect\citeauthoryear{Da Rio et al.}{2012}]{dario12} Da Rio N., Robberto M., 
Hillenbrand L. A., Henning T., Stassun K. G. 2012, ApJ, 748, 14
\bibitem[\protect\citeauthoryear{De Marchi, Paresce \&
    Pulone}{2007}]{demarchi07} De Marchi G., Paresce F., Pulone
  L. 2007, ApJ, 656, L65
\bibitem[\protect\citeauthoryear{De Marchi, Paresce \& Portegies
    Zwart}{2010}]{demarchi10} De Marchi G., Paresce F., Portegies
  Zwart S. 2010, ApJ, 718, 105
\bibitem[\protect\citeauthoryear{De Marchi, Panagia \& Sabbi}{2011}]{demarchi11} De Marchi G., 
Panagia N., Sabbi E. 2011, ApJ, 740, 10
%\bibitem[\protect\citeauthoryear{D'Ercole et al.}{2008}]{dercole08} D'Ercole A.,
%Vesperini E., D'Antona F., McMillan S. L. W., Recchi S. 2008, MNRAS, 391, 825
%\bibitem[\protect\citeauthoryear{Frank \& Gisler}{1976}]{frank76} Frank J., 
%Gisler G. 1976, MNRAS, 176, 533
\bibitem[\protect\citeauthoryear{Elmegreen, Hurst \& Koenig}{2014}]{elmegreen14} Elmegreen B. G., 
Hurst R., Koenig X. 2014, ApJL, 782, L1
\bibitem[\protect\citeauthoryear{Fregeau, Ivanova \& Rasio}{2009}]{fregeau09} Fregeau J. M.,
Ivanova N., Rasio F. A. 2009, ApJ, 707, 1533
\bibitem[\protect\citeauthoryear{Gabor \& Bournaud}{2013}]{gabor13} Gabor J. M., Bournaud F. 2013, 
MNRAS, 434, 606
\bibitem[\protect\citeauthoryear{Gieles, Heggie \& Zhao}{2011}]{gieles11} Gieles M., Heggie D., Zhao H. 2011, MNRAS,413, 2509
\bibitem[\protect\citeauthoryear{Girichidis et al.}{2012a}]{girichidis12a} Girichidis P.,
Federrath C., Banerjee R., Klessen R. S. 2012, MNRAS, 420, 613
\bibitem[\protect\citeauthoryear{Girichidis et al.}{2012b}]{girichidis12b} Girichidis P., 
Federrath C., Allison R., Banerjee R., Klessen R. S. 2012, MNRAS, 420, 3264 
\bibitem[\protect\citeauthoryear{Gratton et al.}{2001}]{gratton01} Gratton R. G., Bonifacio P., 
Bragaglia A., Carretta E., Castellani V., Centurion M., Chieffi A., Claudi R., Clementini G., 
D'Antona F., Desidera S., Francois P., Grundahl F., Lucatello S., Molaro P., Pasquini L., 
Sneden C., Spite F., Straniero O. 2001, A\&A, 369, 87
\bibitem[\protect\citeauthoryear{Gratton, Carretta \& Bragaglia}{2012}]{gratton12}
Gratton R., Carretta E., Bragaglia A. 2012, A\&ARv, 20, 50
\bibitem[\protect\citeauthoryear{Harris}{1996, 2010 update}]{harris96}
  Harris, W. E. 1996, AJ, 112, 1487 (2010 update)
\bibitem[\protect\citeauthoryear{Heggie \& Hut}{2003}]{heggie03}
  Heggie D. C., Hut P. 2003, The Gravitational Million-Body Problem:
  A Multidisciplinary Approach to Star Cluster Dynamics (Cambridge:
  Cambridge University Press)
%\bibitem[\protect\citeauthoryear{Heggie \& Giersz}{2008}]{heggie08} Heggie D. C.,
%Giersz M. 2008, MNRAS, 389, 1858
%\bibitem[\protect\citeauthoryear{Heggie \& Giersz}{2009}]{heggie09} Heggie D. C.,
%  Giersz M. 2009, MNRAS, 397, 46
\bibitem[\protect\citeauthoryear{Henon}{1960}]{henon60} Henon M. 1960,
  Annales d'Astrophysique, 23, 668
\bibitem[\protect\citeauthoryear{Henon}{1969}]{henon69} Henon M. 1969, 
A\&A, 2, 151
\bibitem[\protect\citeauthoryear{Henon}{1973}]{henon73} Henon
  M. 1973, Dynamical Structure and Evolution of Dense Stellar Systems,
  ed. L. Martinet \& M. Mayor (Geneva Obs.)
Pringle J. E. 2006, MNRAS, 373, L90
\bibitem[\protect\citeauthoryear{Hopkins \& Quataert}{2011}]{hopkins11} Hopkins P. F., 
Quataert E. 2011, MNRAS, 415, 1027
\bibitem[\protect\citeauthoryear{Hoyle \& Lyttleton}{1939}]{hoyle39} Hoyle F., 
Lyttleton R. A. 1939, in Proceedings of the Cambridge Philosophical Society, 35, 405
\bibitem[\protect\citeauthoryear{Kirk \& Myers}{2011}]{kirk11} Kirk H., Myers P. C. 2011,
ApJ, 727, 64
\bibitem[\protect\citeauthoryear{Kirk, Offner \& Redmond}{2014}]{kirk14} Kirk H., Offner S. S. R., 
Redmond K. J. 2014, MNRAS, accepted
%\bibitem[\protect\citeauthoryear{Kirk \& Myers}{2012}]{kirk12} Kirk H., Myers P. C. 2012,
%ApJ, 745, 131
\bibitem[\protect\citeauthoryear{Kormendy \& Ho}{2013}]{kormendy13} Kormendy J., 
Ho L. C. 2013, ARA\&A, 51, 511
%\bibitem[\protect\citeauthoryear{Krause et al.}{2012}]{krause12} Krause M., Charbonnel C.,
%Decressin T., Meynet G., Prantzos N., Diehl R. 2012, A\&A, 546, L5
%\bibitem[\protect\citeauthoryear{Krause et al.}{2013}]{krause13} Krause M., Charbonnel C.,
%Decressin T., Meynet G., Prantzos N.. 2013, A\&A, accepted (arXiv:1302.2494)
\bibitem[\protect\citeauthoryear{Krumholz, McKee \& Klein}{2004}]{krumholz04} Krumholz M. R.,
McKee C. F., Klein R. I. 2004, ApJ, 611, 399
\bibitem[\protect\citeauthoryear{Krumholz, McKee \& Klein}{2005a}]{krumholz05a} Krumholz M. R.,
McKee C. F., Klein R. I. 2005, ApJ, 618, 757
\bibitem[\protect\citeauthoryear{Krumholz, McKee \& Klein}{2005b}]{krumholz05b} Krumholz M. R.,
McKee C. F., Klein R. I. 2005, Nature, 438, 332
\bibitem[\protect\citeauthoryear{Krumholz, McKee \& Klein}{2006}]{krumholz06} Krumholz M. R.,
McKee C. F., Klein R. I. 2006, ApJ, 638, 369
\bibitem[\protect\citeauthoryear{Krumholz, Klein \& McKee}{2011}]{krumholz11a} Krumholz M. R.,
Klein R. I., McKee C. F. 2011, ApJ, 740, 74
\bibitem[\protect\citeauthoryear{Krumholz}{2011}]{krumholz11b} Krumholz M. R. 2011, ApJ, 743, 110
\bibitem[\protect\citeauthoryear{Krumholz, Klein \& McKee}{2012}]{krumholz12} Krumholz M. R.,
Klein R. I., McKee C. F. 2012, ApJ, 754, 71
\bibitem[\protect\citeauthoryear{Kryukova et al.}{2012}]{kryukova12} Kryukova E., Megeath S. T., 
Gutermuth R. A., Pipher J., Allen T. S., Allen L. E., Myers P. C., Muzerolle J. 2012, AJ, 144, 31 
\bibitem[\protect\citeauthoryear{Lada \& Lada}{2003}]{lada03} Lada C. J., Lada E. A. 
2003, ARA\&A, 41, 57
\bibitem[\protect\citeauthoryear{Launhardt, Zylka \& Mezger}{2002}]{launhardt02} Launhardt R., 
Zylka R., Mezger P. G. 2002, A\&A, 384, 112
\bibitem[\protect\citeauthoryear{Lee \& Stahler}{2011}]{lee11} Lee A. T., Stahler S. W. 
2011, MNRAS, 416, 3177
\bibitem[\protect\citeauthoryear{Lee \& Stahler}{2013}]{lee13} Lee A. T., Stahler S. W.
2013, A\&A, accepted
\bibitem[\protect\citeauthoryear{Lee et al.}{2014}]{lee14} Lee A. T., Cunningham A. J., McKee C. F.,
Klein R. I. 2014, ApJ, 783, 50
\bibitem[\protect\citeauthoryear{Leigh et al.}{2013a}]{leigh13a} Leigh N. W.,
B\"{o}ker T., Maccarone T. J., Perets H. B. 2013, MNRAS, 429, 2997
\bibitem[\protect\citeauthoryear{Leigh et al.}{2013b}]{leigh13b} Leigh N. W., Giersz M., Webb J. J., 
Hypki A., De Marchi G., Kroupa P., Sills A. 2013, MNRAS, 436, 3399
\bibitem[\protect\citeauthoryear{Lynden-Bell}{1962a}]{lynden-bell62a} Lynden-Bell D. 1962a,
MNRAS, 123, 447
\bibitem[\protect\citeauthoryear{Lynden-Bell}{1962b}]{lynden-bell62b} Lynden-Bell D. 1962, 
MNRAS, 124, 1
\bibitem[\protect\citeauthoryear{Lynden-Bell}{1962c}]{lynden-bell62c} Lynden-Bell D. 1962,
MNRAS, 124, 95
\bibitem[\protect\citeauthoryear{Maccarone \& Zurek}{2012}]{maccarone12}
Maccarone T. J., Zurek D. R. 2012, MNRAS, 423, 2
%\bibitem[\protect\citeauthoryear{Maeder}{2009}]{maeder09} Maeder A. 2009,
%  Physics, Formation and Evolution of Rotating Stars. Berlin: Springer-Verlag
\bibitem[\protect\citeauthoryear{Marks, Kroupa \& Baumgardt}{2008}]{marks08} Marks M.,
Kroupa P., Baumgardt H. 2008, MNRAS, 386, 2047
%\bibitem[\protect\citeauthoryear{Marks \& Kroupa}{2010}]{marks10} Marks M.,
%Kroupa P. 2010, MNRAS, 406, 2000
\bibitem[\protect\citeauthoryear{Maschberger et al.}{2010}]{maschberger10} Maschberger Th., 
Clarke C. J., Bonnell I. A., Kroupa P. 2010, MNRAS, 404, 1061 
\bibitem[\protect\citeauthoryear{McKee \& Ostriker}{2007}]{mckee07} McKee C. F.,
Ostriker E. C. 2007, ARA\&A, 45, 565
\bibitem[\protect\citeauthoryear{Merritt}{2013}]{merritt13} Merritt D. 2013, Dynamics and Evolution
of Galactic Nuclei (Princeton:  Princeton University Press)
\bibitem[\protect\citeauthoryear{Moeckel \& Bonnell}{2009a}]{moeckel09a} Moeckel N., 
Bonnell I. A. 2009, MNRAS, 396, 1864
\bibitem[\protect\citeauthoryear{Moeckel \& Bonnell}{2009b}]{moeckel09b} Moeckel N.,
Bonnell I. A. 2009, MNRAS, 400, 657
\bibitem[\protect\citeauthoryear{Moeckel et al.}{2012}]{moeckel12} Moeckel N., Holland C., 
Clarke C. J., Bonnell I. A. 2012, MNRAS, 425, 450
\bibitem[\protect\citeauthoryear{Odenkirchen et al.}{2003}]{odenkirchen03} Odenkirchen M., 
Grebel E. K., Dehnen W., Rix H.-W., Yanny B., Newberg H. J., Rockosi C. M., Martinez-Delgado D., 
Brinkmann J., Pier J. R. 2003, AJ, 126, 2385 
\bibitem[\protect\citeauthoryear{Offner, Hansen \& Krumholz}{2009}]{offner09a} Offner S. S. R., 
Hansen C. E., Krumholz M. R. 2009, ApJL, 704, L124
\bibitem[\protect\citeauthoryear{Offner et al.}{2009}]{offner09b} Offner S. S. R., Klein R. I., 
McKee C. F., Krumholz M. R. 2009, ApJ, 703, 131
\bibitem[\protect\citeauthoryear{Offner \& McKee}{2011}]{offner11} Offner S. S. R., 
McKee C. F. 2011, ApJ, 736, 53
\bibitem[\protect\citeauthoryear{Osborn}{1971}]{osborn71} Osborn W. 1971, Observatory, 91, 223
\bibitem[\protect\citeauthoryear{Ostriker}{1999}]{ostriker99} Ostriker E. C. 1999, 
ApJ, 513, 252
%\bibitem[\protect\citeauthoryear{Park \& Ricotti}{2013}]{park13} Park K., Ricotti M. 2013, 
%ApJ, submitted (arXiv:1211.0542)
\bibitem[\protect\citeauthoryear{Parker \& Meyer}{2012}]{parker12} Parker R. J., Meyer M. R. 
2012, MNRAS, 427, 637
\bibitem[\protect\citeauthoryear{Parker et al.}{2014}]{parker14} Parker R. J., Wright N. J., 
Goodwin S. P., Meyer M. R. 2014, MNRAS, 438, 620
\bibitem[\protect\citeauthoryear{Perets, Hopman \& Alexander}{2007}]{perets07} Perets H. B., 
Hopman C., Alexander T. 2007, ApJ, 656, 709
\bibitem[\protect\citeauthoryear{Phan-Bao et al.}{2011}]{phan-bao11} Phan-Bao N., Lee C.-F., 
Ho P. T. P., Tang Y.-W. 2011, ApJ, 735, 14
%\bibitem[\protect\citeauthoryear{Phinney \& Sigurdsson}{1991}]{phinney91} Phinney S. E.,
%Sigurdsson S. 1991, Nature, 349, 220
\bibitem[\protect\citeauthoryear{Piotto et al.}{2007}]{piotto07}
Piotto G., Bedin L. R., Anderson J., King I. R., Cassisi S., Milone A. P., Villanova S., Pietrin-
ferni A., Renzini A. 2007, ApJ, 661, L53
\bibitem[\protect\citeauthoryear{Portegies Zwart et al.}{2004}]{portegieszwart04} Portegies Zwart S. F., 
Baumgardt H., Hut P., Makino J., McMillan S. L. W. 2004, Nature, 428, 724
%\bibitem[\protect\citeauthoryear{Priestley, Ruffert \& Salaris}{2011}]{priestley11} Priestley W., 
%Ruffert M., Salaris M. 2011, MNRAS, 411, 1935
\bibitem[\protect\citeauthoryear{Rephaeli \& Salpeter}{1980}]{rephaeli80} Rephaeli Y., 
Salpeter E. E. 1980, ApJ, 240, 20
\bibitem[\protect\citeauthoryear{Ruderman \& Spiegel}{1971}]{ruderman71} Ruderman M. A., 
Spiegel E. A. 1971, ApJ, 165, 1
\bibitem[\protect\citeauthoryear{Ruffert}{1997}]{ruffert97} Ruffert M. 1997, A\&A, 317, 793
%\bibitem[\protect\citeauthoryear{Rybicki \& Lightman}{1979}]{rybicki79} Rybicki G. B.,
%Lightman A. P. 1979, Radiative Processes in Astrophysics (New York: Wiley-Interscience)
\bibitem[\protect\citeauthoryear{Schinnerer, B\"oker \& Meier}{2003}]{schinnerer03} Schinnerer E.,
  B\"oker T., Meier D. S. 2003, ApJ, 591, L115
\bibitem[\protect\citeauthoryear{Schinnerer et al.}{2006}]{schinnerer06} Schinnerer E., 
B\"oker T., Emsellem E., Lisenfeld U. 2006, AJ, 649, 181 
\bibitem[\protect\citeauthoryear{Spitzer}{1969}]{spitzer69} Spitzer L. Jr. 1969, ApJ, 
158, 139
\bibitem[\protect\citeauthoryear{Spitzer}{1987}]{spitzer87} Spitzer L. Jr. 1987,
Dynamical Evolution of Globular Clusters (Princeton, NJ: Princeton Univ. Press)
\bibitem[\protect\citeauthoryear{Strader et al.}{2012}]{strader12} Strader J., Chomiuk L.,
Maccarone T. J., Miller-Jones J. C. A., Seth A. C. 2012, Nature, 490, 71
%\bibitem[\protect\citeauthoryear{Taylor \& Wood}{1975}]{taylor75} Taylor R. J., 
%Wood P. R. 1975, MNRAS, 171, 467
\bibitem[\protect\citeauthoryear{Tremaine, Ostriker \& Spitzer}{1975}]{tremaine75}
Tremaine S. D., Ostriker J. P., Spitzer L. Jr. 1975, ApJ, 196, 407
\bibitem[\protect\citeauthoryear{Trenti \& van der Marel}{2013}]{trenti13} Trenti M., 
van der Marel R. 2013, MNRAS, doi: 10.1093/mnras/stt1521
%\bibitem[\protect\citeauthoryear{Tutukov}{1978}]{tutukov78} Tutukov A. V.
%1978, A\&A, 70, 57
%\bibitem[\protect\citeauthoryear{Vesperini \& Heggie}{1997}]{vesperini97}
%Vesperini E., Heggie D. C. 1997, MNRAS, 289, 898
\bibitem[\protect\citeauthoryear{Vishniac}{1978}]{vishniac78} Vishniac E. T. 1978, ApJ, 223, 986
\bibitem[\protect\citeauthoryear{von Hippel \&
    Sarajedini}{1998}]{vonhippel98} von Hippel T., Sarajedini A. 1998,
  AJ, 116, 1789
\bibitem[\protect\citeauthoryear{Wang et al.}{2011}]{wang11} Wang W., Boudreault S., Goldman B., 
Henning T., Caballero J. A., Bailer-Jones C. A. L. 2011, A\&A, 531, 164
%\bibitem[\protect\citeauthoryear{Webb, Harris \& Sills}{2012}]{webb12} Webb J. J., 
%Harris W. E., Sills A. 2012, ApJ, 759, 39

\end{thebibliography}
\end{document}